\documentclass[11pt]{article}
\pdfoutput=1
\usepackage{jheppub}
\usepackage{graphicx}
\usepackage{amssymb}
\usepackage{amsmath,amssymb}
\usepackage{slashed}
\usepackage{hyperref}
\usepackage{caption}
\usepackage{xcolor}
\usepackage{dsfont}
\usepackage{verbatim}
\usepackage{subfig}

\newcommand\beq{\begin{equation}}
\newcommand\eeq{\end{equation}}

\title{New and Old Fermionic Dualities from 3d Bosonization}

\preprint{\today}

\author[a]{Kyle Aitken,}
\author[b]{ Changha Choi,}
\author[a]{ Andreas Karch}
\affiliation[a]{Department of Physics, University of Washington, Seattle, WA, 98195-1560, USA}
\affiliation[b]{Physics and Astronomy Department,
Stony Brook University, Stony Brook, NY 11794, USA}

\abstract{We construct novel fermion-fermion dualities in $2+1$-dimensions using 3d bosonization dualities. This is achieved by relating two-node quiver theories using both the flavor-bounded and flavor-violated 3d bosonization dualities. Such quivers can be viewed as a generalization of the fermionic particle-vortex duality. A special case of these quivers exhibits a $\mathbb{Z}_2$ symmetry under interchange of the two nodes. Using orbifold techniques, we show that such dualities provide a novel way of deriving known 3d bosonization dualities with adjoint matter, thus unifying the non-Abelian bosonization dualities in an even larger duality web. We then use this construction to derive new dualities involving adjoint matter.
}

\begin{document}
\maketitle

\section{Introduction}

In the past few years, a family of new non-supersymmetric dualities between Chern-Simons-matter theories in $2+1$-dimensions has been conjectured \cite{Aharony:2015mjs,Hsin:2016blu,Jensen:2017bjo,Benini:2017aed} involving fundamental or antifundamental representations, i.e. rank-one matter. Schematically, these dualities are \cite{Aharony:2015mjs}
\begin{subequations}
\label{eq:aharony}
\begin{align}
U(k)_{N}\text{ with \ensuremath{N_{f}} \ensuremath{\Phi}}\qquad & \leftrightarrow\qquad SU(N)_{-k+N_{f}/2}\text{ with \ensuremath{N_{f}} \ensuremath{\psi}}\label{eq:aharony 1}\\
SU(k)_{N}\text{ with \ensuremath{N_{f}} \ensuremath{\phi}}\qquad & \leftrightarrow\qquad U(N)_{-k+N_{f}/2}\text{ with \ensuremath{N_{f}} \ensuremath{\Psi}}\label{eq:aharony 2}
\end{align}
\end{subequations}
where $\phi, \Phi$ are fundamental scalars; $\psi, \Psi$ are fundamental fermions; and throughout this work we use ``$\leftrightarrow$'' to denote a duality conjectured to hold in the IR limit. In the simplest ``flavor-bounded'' regime the number of matter fields is constrained so that $N_f\leq k$ in the above expressions.

More recently, various extensions of the dualities of \eqref{eq:aharony} have been proposed.  There is evidence the duality continues to hold in the ``flavor-violated'' regime where $k < N_f < N_*(k,N)$, for some yet-unknown function $N_*$ \cite{Komargodski:2017keh}. Additionally, a form of the duality which has both rank-one fermionic and bosonic matter on each side has also been proposed \cite{Jensen:2017bjo,Benini:2017aed}. Since the dualities of \eqref{eq:aharony} are a special case of this more general duality, we will refer to it as the ``master duality''. Finally, there also exists conjectures for dualities very much analogous to \eqref{eq:aharony}, but instead involving adjoint, symmetric, or antisymmetric representations, i.e. rank-two matter \cite{Gomis:2017ixy, Choi:2018tuh}.

\begin{figure}
\begin{centering}
\includegraphics[scale=0.5]{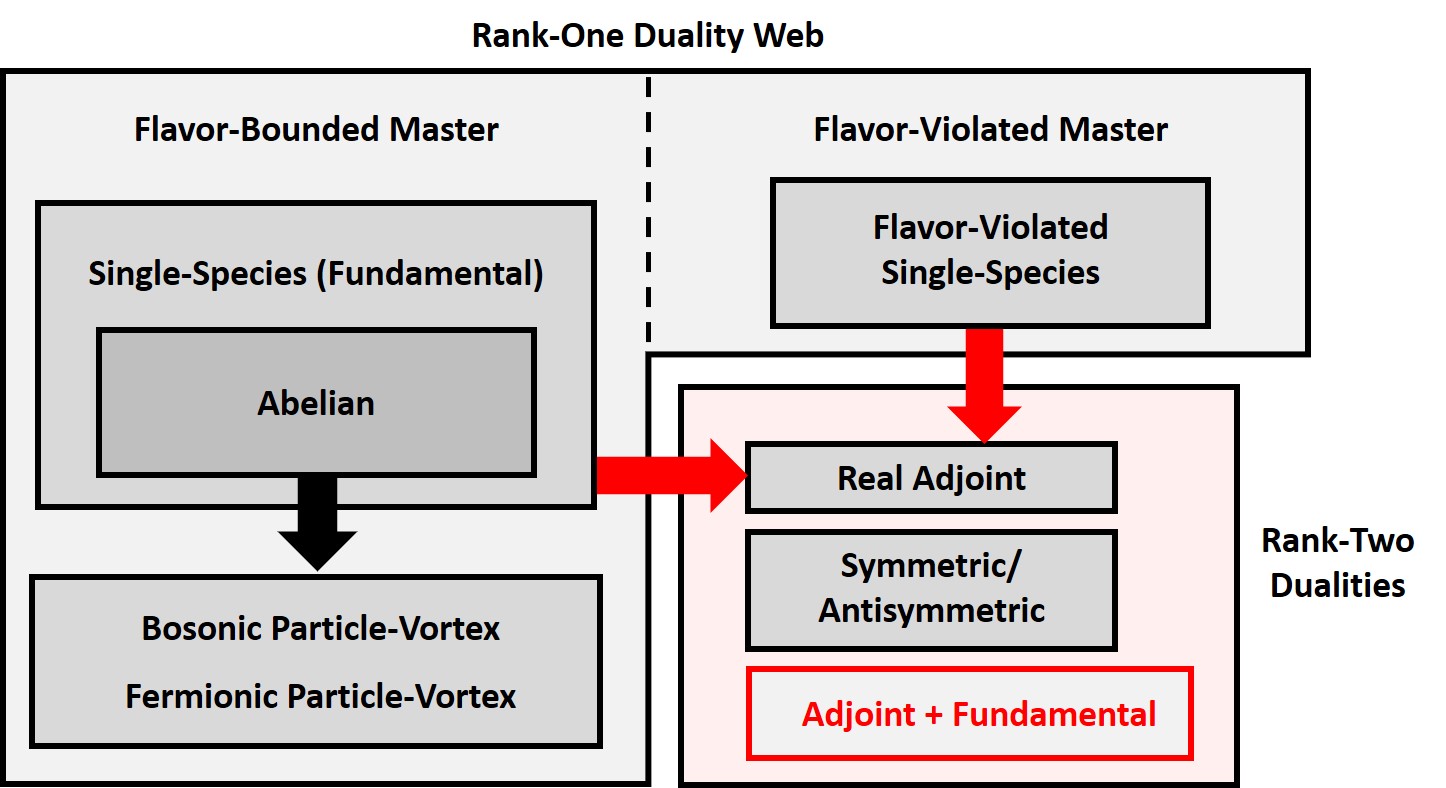}
\par\end{centering}
\caption{Web of bosonization dualities in $2+1$-dimensions. The contributions of this work are shaded in red. Specifically, we show that the rank-two adjoint matter dualities can be connected to the rest of the web of dualities (red arrows). We also conjecture several new rank-two dualities including a duality with both fundamental and adjoint representation matter. \label{fig:Web-of-bosonizatino 2r}}
\end{figure}

Using the master duality, the single-species dualities of \eqref{eq:aharony}, and the Abelian limit of \eqref{eq:aharony}, a large web of rank-one dualities has been constructed, see Fig. \ref{fig:Web-of-bosonizatino 2r} \cite{Karch:2016sxi, Seiberg:2016gmd, Jensen:2017dso, Aitken:2018cvh, Aitken:2019mtq}. Although the rank-two dualities bear a strong resemblance to the rank-one 3d bosonization dualities, they have so far remained disconnected from aforementioned web.

The usual duality web construction involving adding a background term/decoupled theory and gauging global symmetries was explored within the context of rank-one dualities in the work of \cite{Jensen:2017dso,Aitken:2018cvh, Aitken:2019mtq} to construct \emph{bosonic} quivers. That is, by gauging non-Abelian global symmetries the rank-one dualities can be used to derive dualities between theories with product gauge groups coupled to each other with bifundamental scalar matter. Such quiver dualities were generalized to an arbitrary number of nodes. The two-node case was shown to be a generalization of the bosonic particle-vortex duality \cite{Peskin:1977kp, Dasgupta:1981zz}. A special case of the general quiver was shown to have application for dualities of interfaces in QCD$_{4}$ \cite{Gaiotto:2017tne}.

In this paper, we follow a similar methodology but instead construct two-node quiver dualities which have \emph{fermionic} matter on both ends. We will start by constructing said dualities using only the flavor-bounded 3d bosonization dualities. As one might expect from previous work \cite{Aitken:2018cvh}, such quivers provide a generalization of Son's fermionic particle-vortex duality \cite{Son:2015xqa}. Nevertheless, such quivers are much more strongly constrained compared to their bosonic counterparts as already noted in Ref. \cite{Jensen:2017dso}.

This motivates us to consider the construction of such two-node dualities using the flavor-violated 3d bosonization duality. This relaxes constraints on parameters at the cost of a more complicated phase diagram. A special case of these flavor-violated two-node quivers exhibits a $\mathbb{Z}_2$ symmetry under interchange of its two nodes.

Our main focus of this paper will be a novel procedure of deriving new and old fermionic dualities using this  $\mathbb{Z}_2$-symmetric quiver. In particular, we combine said quivers with an orbifolding procedure, which allows us to connect the rank-one 3d bosonization duality to dualities with rank-two matter, see Fig. \ref{fig:Web-of-bosonizatino 2r}.  This links the rank-two dualities to the rank-one bosonization dualities \cite{Karch:2016sxi, Seiberg:2016gmd}, meaning the web of 3d bosonization dualities is larger than previously thought.

After demonstrating we can recover the aforementioned rank-two dualities, we derive new dualities involving rank-two matter. The first of these closely resembles the known rank-two dualities with the $U(1)$ factors shuffled around. We also explore the possibility of adding fundamental matter on each of the nodes, which gives rise to a new duality involving both adjoint and fundamental matter.

The paper is outlined as follows. In Sec. \ref{sec:review} we briefly review the 3d bosonization dualities with rank-one matter. Next, in Sec. \ref{sec:two-node} we first construct the flavor-bounded two-node quiver, and then the flavor-violated version. Sec. \ref{sec:orbs} introduces the orbifolding procedure to derive new and old fermionic dualities. Finally, we discuss possible extensions of our work and conclude in the Sec. \ref{sec:conclusion}.

Appendix \ref{app:two-node} contains the details of the two-node derivations. Appendix \ref{app:ferm_pv} discusses how the flavor-bounded two-node is connected to the fermionic particle-vortex duality of Ref. \cite{Son:2015xqa}. In Appendix \ref{app:so/sp}, we summarize the generalization of the flavor-violated fermionic quiver dualities for orthogonal and symplectic gauge groups. Finally, we illustrate the gravitational counterterm matching as a consistency check for the two-node quiver dualities in Appendix \ref{app:gravitational}.

\section{Review of 3d Bosonization Dualities}
\label{sec:review}

The duality in \eqref{eq:aharony 1} conjectures an IR equivalence between the following two Lagrangians,\footnote{Since the fermion suffers from the parity anomaly in $2+1$-dimensions \cite{Redlich:1983kn, Niemi:1983rq}, implicit in these dualities are Pauli-Villars regulators. These take the form of $\eta$-invariants. It is convention in the literature to neglect such terms in the Lagrangian, which amounts to the rewriting
\begin{align}
i\bar{\psi}\slashed{D}_c \psi -i\left[-\frac{1}{2}\frac{N_f}{4\pi}\text{Tr}_k\left(cdc - i\frac{2}{3}c^3\right)\right] \quad \to \quad i\bar{\psi}\slashed{D}_c \psi.
\end{align}
We will follow this convention when writing Lagrangians, but will explicitly denote such $\eta$-invariant factors otherwise.
}
\begin{subequations}
\label{eq:aharony lag 1}
\begin{align}
\mathcal{L}_{SU} & =i\bar{\psi}\slashed{D}_{b^{\prime}-C+\tilde{A}_{1}}\psi-i\left[\frac{N_{f}-k}{4\pi}\text{Tr}_{N}\left(b^{\prime}db^{\prime}-i\frac{2}{3}b^{\prime3}\right)\right] \nonumber \\ &-i\left[\frac{N}{4\pi}\text{Tr}_{N_{f}}\left(CdC-i\frac{2}{3}C^{3}\right)+\frac{N\left(N_{f}-k\right)}{4\pi}\tilde{A}_{1}d\tilde{A}_{1}\right],\\
\mathcal{L}_{U} & =\left|D_{c-C}\Phi\right|^{2}-i\left[\frac{N}{4\pi}\text{Tr}_{k}\left(cdc-i\frac{2}{3}c^{3}\right)-\frac{N}{2\pi}\text{Tr}_{k}\left(c\right)d\tilde{A}_{1}\right],\label{eq:aharony 2 lag 2r}
\end{align}
\end{subequations}
which is subject to the flavor bounds $k\geq N_{f}$.  The Lagrangians contain Chern-Simons terms for both dynamical and background gauge fields. Table \ref{tab:Definitions-of-various ferm quiv-1} shows our definitions of fields. Uppercase letters are used for background fields and lowercase for dynamical fields.

\begin{table}
\begin{centering}
\begin{tabular}{|c|c|c||c|c|}
\cline{2-5}
\multicolumn{1}{c|}{} & \multicolumn{2}{c||}{\textbf{Gauge Fields}} & \multicolumn{2}{c|}{\textbf{Background Fields}}\tabularnewline
\hline
\textbf{Symmetry} & $SU\left(N\right)$ & $U\left(k\right)$ & $SU\left(N_{f}\right)$ & $U\left(1\right)_{m,b}$\tabularnewline
\hline
\textbf{Field} & $b_{\mu}^{\prime}$ & $c_{\mu}$  & $C_{\mu}$ & $\tilde{A}_{1\mu}$\tabularnewline
\hline
\end{tabular}
\par\end{centering}
\caption{Definitions of various gauge fields used in the duality of \eqref{eq:aharony lag 1}.\label{tab:Definitions-of-various ferm quiv-1}}
\end{table}

We can simultaneously deform both sides of the above duality by adding mass terms for the respective matter. For large enough mass, we can integrate out the matter leaving behind a topological field theory (TFT). The mass mapping $m_{\psi}\leftrightarrow-m_{\Phi}^{2}$ gives TFTs which are level-rank dual to one another,
\begin{align}
\label{eq:lr 2r}
SU(N)_{\pm k}\qquad & \Leftrightarrow\qquad U(k)_{\mp N}.
\end{align}
This duality has been rigorously proven to hold at all scales, and is part of a larger class of level-rank dualities, which also includes \cite{Hsin:2016blu}
\begin{align}
U(N)_{k,k\pm N}\qquad & \Leftrightarrow\qquad U(k)_{-N,-N\mp k}.\label{eq:lr 2r u u}
\end{align}
Here we are using the notation
\begin{equation}
U(N)_{P,Q}=\frac{SU(N)_{P}\times U(1)_{NQ}}{\mathbb{Z}_{N}}
\end{equation}
to denote Chern-Simons terms where the Abelian and non-Abelian parts have different levels. We use the shorthand $U(k)_{N,N}\equiv U(k)_N$ when the levels are equal, up to rank dependence.

In this work it will be important to be careful about the representation of matter fields under various gauge groups, both dynamical and background. Our notation for covariant derivatives is such that
\begin{subequations}
\begin{align}
\left(D_{b^{\prime}-C+\tilde{A}_{1}}\right)_{\mu}\psi & =\left[\partial_{\mu}-i\left(b_{\mu}^{\prime}\mathds{1}_{N_{f}}+C_{\mu}\mathds{1}_{N}+\tilde{A}_{1\mu}\mathds{1}_{NN_{f}}\right)\right]\psi,\\
\left(D_{c-C}\right)_{\mu}\Phi & =\left[\partial_{\mu}-i\left(c_{\mu}\mathds{1}_{N_{f}}+C_{\mu}\mathds{1}_{k}\right)\right]\Phi,
\end{align}
\end{subequations}
where $b_{\mu}^{\prime}$ and $c_{\mu}$ are fundamental representations and $C_{\mu}$ is in the anti-fundamental representation of their respective gauge groups. $\mathds{1}_N$ is the $N\times N$ identity matrix. That is, antifundamental representations will be denoted by a minus sign in the subscript of the covariant derivative.

As mentioned in the introduction, an extension of the 3d bosonization duality to $k < N_f < N_*$ was proposed in Ref. \cite{Komargodski:2017keh}. On the fermion end, the phase diagram as a function of mass now has an intermediate ``quantum region'' bordered by two second-order phase transitions at $m_\psi = \pm m_*$.\footnote{In general, the phase transition points do not necessarily need to exist at the same mass magnitude $m_*$. However, one can tune the bare mass parameter to make two transition points symmetrical, and we will assume that has been done throughout this work.} At both transition points there is a bosonized dual description in terms of $N_f$ scalars in the fundamental representation,
\begin{equation}
\label{eq:aharony fv}
SU(N)_{-k+N_f/2}\text{ with \ensuremath{N_f\;\psi}}\qquad\leftrightarrow\qquad\begin{cases}
U(k)_{N}\text{ with  \ensuremath{N_f\;\Phi_1}} & m_{\psi}=-m_{*} \\
U(N_f-k)_{-N}\text{ with \ensuremath{N_f\;\Phi_2}} & m_{\psi}=m_{*}.
\end{cases}
\end{equation}
We will sometimes label the bosonized duals by whether they are dual to the ``$m_\psi>0$'' or ``$m_\psi<0$'' region of the fermionic phase diagram (with appropriate mass deformations). The quantum phase cannot be described by the weakly coupled analysis from the UV fermionic theory. The two mutually non-local scalar dual theories describe the same intermediate quantum phase for $m_\psi \in (-m_*,m_*) $ in terms of non-linear $\sigma$-model with some Wess-Zumino term. The structure of this phase diagram will be the starting point for constructing the phase diagram of flavor-violated two-node fermionic quivers in Sec. \ref{subsec:Flavor-Violated 2r}.

\section{Two-Node Fermionic Quivers}
\label{sec:two-node}

It has been shown in Refs. \cite{Aitken:2018cvh, Aitken:2019mtq} that we can use two-node quivers to create generalizations of the bosonic particle-vortex duality. In this section we construct the fermionic equivalent. Flavor constraints will imposes tighter bounds on the possible theories we can create, since the intermediate quivers contain scalars. This will motivate us to also consider two-node fermionic quivers which use the flavor-violated version of 3d bosonization. In this section we will mostly stick to schematic notation for brevity. Details of this derivation at the Lagrangian level can be found in Appendix \ref{app:two-node}.

\subsection{Flavor-Bounded}
\label{sec:flavor-bounded}

The general procedure for constructing two-node quiver dualities involves promoting the $SU(N_f)$ and/or $U(1)$ global symmetries after adding appropriate background Chern-Simons terms \cite{Jensen:2017dso}. For example, promoting the background flavor symmetries to be dynamical, the schematic form of the 3d bosonization dualities in \eqref{eq:aharony} is given by
\begin{subequations}
\begin{align}
U(k)_{N}\times SU(N_f)_{0} +\Phi &\quad\leftrightarrow\quad SU(N)_{-k+N_f/2}\times SU(N_f)_{N/2}+\psi,\label{eq:fq aharony 2} \\
SU(k)_{N}\times U(N_{f})_{0}+\phi &\quad\leftrightarrow\quad U(N)_{-k+N_{f}/2}\times U(N_f)_{N/2} + \Psi. \label{eq:fq aharony 1}
\end{align}
\end{subequations}
Each side either has a single bifundamentally charged fermion or boson, e.g. the right-hand side of \eqref{eq:fq aharony 1} has fermion in the $(\mathbf{N}, \mathbf{\bar{N}}_f)$ representation.  In the latter of these dualities we have also promoted the $U(1)$ global symmetry. When we do this we introduce a new $U(1)$ global symmetry whose corresponding background field couples to the promoted $\tilde{A}_{1}$ via a BF term.

Relabeling parameters and adding background terms (before promotion), the duality \eqref{eq:fq aharony 2} conjectures the following two theories are dual
\begin{align}
\text{Theory A:}\qquad & SU(N_{1})_{-k_{1}+N_{2}/2}\times SU(N_{2})_{-k_{2}+N_{1}/2} + \psi, \label{eq:theory a}\\
\text{Theory B':}\qquad & U(k_1)_{N_1}\times SU(N_2)_{-k_2} + \Phi,\label{eq:theory b 2r}
\end{align}
which is subject to the flavor bound $k_{1}\geq N_{2}$. Now also rewrite the second duality, \eqref{eq:fq aharony 1}, again with a special choice of added background terms and relabeling,
\begin{align}
\text{Theory B'':}\qquad & U(k_1)_{N_1}\times SU(N_2)_{-k_2} + \phi, \\
\text{Theory C:}\qquad & U(k_1)_{N_{1}-k_{2}/2}\times U(k_2)_{N_{2}-k_{1}/2}+\Psi, \label{eq:theory c 2r}
\end{align}
subject to the flavor bound $N_{2}\geq k_{1}$.

\begin{figure}
\begin{centering}
\includegraphics[scale=0.65]{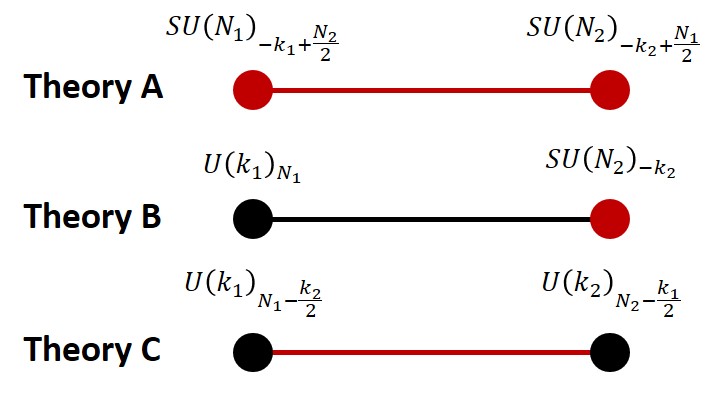}
\par\end{centering}
\caption{Generalized fermion quiver duality of \eqref{eq:flaovr-bounded lag} with flavor-bounded dualities. Theory B represents both Theories $\text{B'}$ and $\text{B''}$ since they are identical. Our quiver notation is such that we use filled red (black) nodes to represent dynamical $SU$ ($U$) gauge groups. The red (black) links between said circles represent fermions (scalars) charged under the corresponding gauge groups. This duality is subject to the flavor bound $N_2 = k_1$.
\label{fig:Generalized-fermion-quiver}}
\end{figure}

Theories $\text{B'}$ and $\text{B''}$ are identical, so we will collectively call them Theory $\text{B}$. Since Theory A and Theory C are both dual to Theory B, they must also be dual to one another. That is, we have a fermion-fermion duality between \eqref{eq:theory a} and \eqref{eq:theory c 2r}. Each side has a bifundamental fermion coupled to two gauge fields with Chern-Simons terms. Note simultaneously satisfying both of the flavor bounds requires $N_{2}=k_{1}$, but $N_{1}$ and $k_{2}$ are unbounded. At the Lagrangian level, this is a duality between the theories
\begin{subequations}
\label{eq:flaovr-bounded lag}
\begin{align}
\mathcal{L}_{\text{A}} & =i\bar{\psi}\slashed{D}_{b^{\prime}-c^{\prime}+\tilde{A}_{1}}\psi-i\left[\frac{N_{2}-k_{1}}{4\pi}\text{Tr}_{N_{1}}\left(b^{\prime}db^{\prime}-i\frac{2}{3}b^{\prime3}\right)\right]\nonumber \\
 & -i\left[\frac{N_{1}-k_{2}}{4\pi}\text{Tr}_{N_{2}}\left(c^{\prime}dc^{\prime}-i\frac{2}{3}c^{\prime3}\right)+\frac{N_{1}N_{2}}{4\pi}\tilde{A}_{1}d\tilde{A}_{1}\right],\label{eq:theory a bounded 2r}\\
\mathcal{L}_{\text{C}} & =i\bar{\Psi}\slashed{D}_{-c+g+\tilde{A}_{1}}\Psi-i\left[\frac{N_{1}}{4\pi}\text{Tr}_{k_{1}}\left(gdg-i\frac{2}{3}g^{3}\right)+\frac{N_{2}}{4\pi}\text{Tr}_{k_{2}}\left(cdc-i\frac{2}{3}c^{3}\right)\right],\label{eq:theory c bounded 2r}
\end{align}
\end{subequations}
where $b'\in su(N_1)$, $c'\in su(N_2)$, $g\in u(k_1)$, and $c\in u(k_2)$. The quiver construction is summarized in Fig. \ref{fig:Generalized-fermion-quiver}.

In Appendix \ref{app:ferm_pv}, we will discuss more details about this fermion-fermion duality constructed from the flavor-bounded bosonization duality. This includes showing Son's fermionic particle-vortex duality \cite{Son:2015xqa} is simply the $N_{1}=N_{2}=k_{1}=k_{2}=1$ limit. For now, we move onto deriving a two-node fermionic duality without such tight flavor constraints.

\subsection{Flavor-Violated}
\label{subsec:Flavor-Violated 2r}

\begin{figure}
\begin{centering}
\includegraphics[scale=0.6]{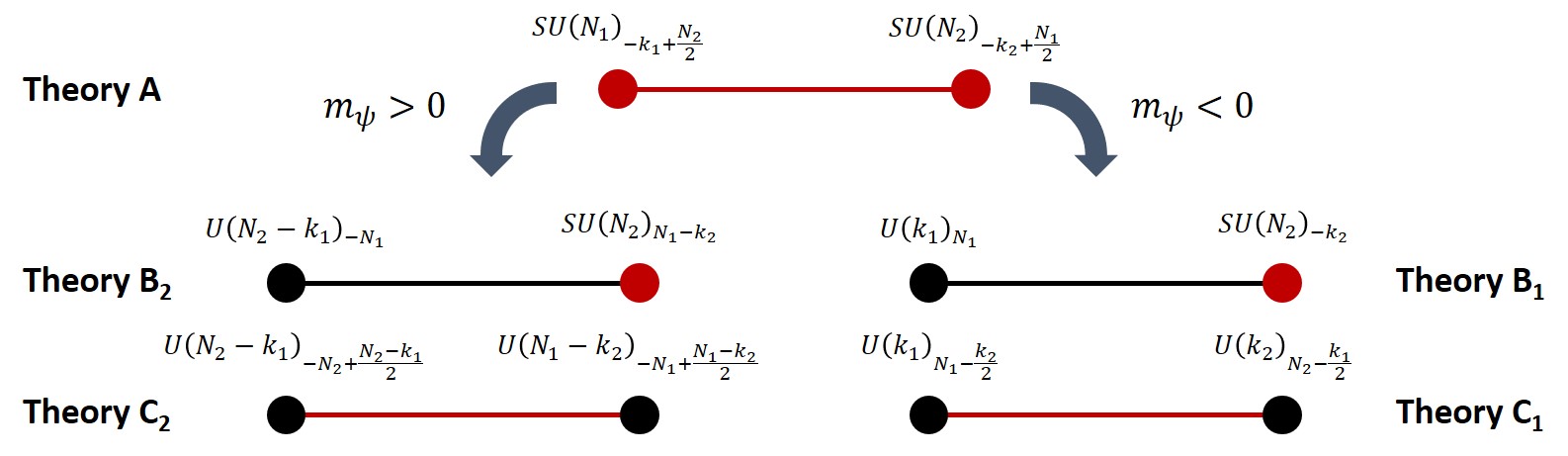}
\par\end{centering}
\caption{Generalized fermion quiver duality with flavor-violated bosonization duality. The first step uses the flavor-violated 3d bosonization duality, while the second step uses the flavor-bounded version. The diagram is valid for the $k_1<N_2 \leq N_*(N_2, k_1),~k_2<N_1 \leq N_*(N_1,k_2)$. \label{fig:two_node_violated 2r}}
\end{figure}

Above, we saw the derivation of the two-node fermions was constrained by the flavor-bound of the 3d bosonization dualities. It is also possible to derive two-node quivers using the flavor-violated duality proposed by Ref. \cite{Komargodski:2017keh}. This comes at the cost of the full phase diagram of Theory B being described by two distinct dual theories. We then apply a flavor-bounded bosonization duality to Theory B, which means Theory C's phase diagram requires two fermionic theories. This construction is summarized in Fig. \ref{fig:two_node_violated 2r}.

In more detail, we start with the same Theory A as we did for the flavor-bounded version above, \eqref{eq:theory a}. We now use the flavor-violated bosonization duality, which lifts the earlier flavor-bound constraint, $k_{1}\geq N_{2}$, to $k_{1}<N_{2}\leq N_{*}(N_{2},k_{1})$ instead. Here, $N_{*}$ is some unknown function of $N$ and $k$ as discussed in Ref. \cite{Komargodski:2017keh}. This yields a Theory B which must be described by two distinct scalar duals. One of these scalar duals, corresponding to the $m_{\psi}<0$ side, is identical to \eqref{eq:theory b 2r}. The $m_{\psi}>0$ side is in general distinct. Schematically, the full phase diagram is described by the theories
\begin{align} \label{eq:2nodebosefermi}
\text{Theory A:}\qquad & SU(N_1)_{-k_{1}+N_{2}/2}\times SU(N_2)_{-k_{2}+N_{1}/2}+\psi,\\
\text{Theory B\ensuremath{_{1}}:}\qquad & U(k_1)_{N_{1}}\times SU(N_2)_{-k_{2}}+\phi_1,\\
\text{Theory B\ensuremath{_{2}}:}\qquad & U(N_{2}-k_{1})_{-N_{1}}\times SU(N_2)_{N_{1}-k_{2}}+\phi_2,
\end{align}
where subscripts $1,2$ correspond to the $m_\psi<0$ and $m_\psi>0$ theories, respectively.

Next, we can use flavor-bounded 3d bosonization to find a fermionic theory which matches the intermediate scalar theory. Since there are two scalar duals in the intermediate theory, we will also need two fermion theories to describe the full phase diagram. Once more, the $m_{\psi}<0$ side is identical to what we found above in \eqref{eq:theory c 2r}. A subtlety is that, due to the flipping of the sign of the level, on the $m_{\psi}>0$ side we must use the time-reversed version of \eqref{eq:fq aharony 2}. This yields the theories
\begin{align}
 \label{eq:2nodefermifermi}
\text{Theory C\ensuremath{_{1}}:} & \qquad U(k_1)_{N_{1}-k_{2}/2}\times U(k_2)_{N_{2}-k_{1}/2}+\Psi_1,\\
\text{Theory C\ensuremath{_{2}}:} & \qquad U(N_{1}-k_{2})_{-N_{2}+\left(N_{2}-k_{1}\right)/2}\times U(N_{2}-k_{1})_{-N_{1}+\left(N_{1}-k_{2}\right)/2}+\Psi_2.
\end{align}
Thus we again arrive at two-node fermionic quiver theories on either end. Once again they are conjectured to be dual to one another, but the flavor bound of $N_2=k_1$  has been replaced by $k_{1}<N_{2}<N_{*}(N_2,k_1)$ and $k_{2}<N_{1}<N_{*}(N_1,k_2)$, which allows us a little more freedom in choosing parameters.\footnote{An important point is that we need further condition $k_{2}<N_{1}\leq N_{*}(N_{1},k_{2})$ to have mutually non-local dual descriptions as the ungauged version. The main reason for this additional constraint is that since we have two dynamical gauge groups, we can equivalently view the first gauge group as the flavor symmetry group and second as originally dynamical gauge group before constructing the quiver.}

The explicit form of the Lagrangians for Theories $\text{B}_{2}$ and $\text{C}_{2}$ are given in \eqref{eq:flavor violated lag}. Note the presence of the mixed BF term between $c$ and $g$, which arises from the fact that dual fermion is charged under both $U(1)$ groups. Since the $m_{\psi}<0$ side is identical to the flavor-bounded case, $\mathcal{L}_{\text{C}_{1}}$ is given by \eqref{eq:theory c bounded 2r}. The BF term between dynamical gauge fields $c$ and $g$ is also present in \eqref{eq:theory c bounded 2r}, but is hidden due to our convention of $\eta$-invariant terms (see footnote 1).

Finally, it is slightly subtle to see how the quantum phases described by Theories C$_1$ and C$_2$ are dual to one another. We denote a level $\pm1$ BF term between two unitary gauge fields $c$ and $g$, e.g. $\pm\frac{1}{2\pi}\text{Tr}_{k_1} (c)d\text{Tr}_{k_2} (g)$, as ``$\pm \text{BF}$''. The intermediate quantum phase is then described by the exact duality,
\begin{align} \label{eq:levelrankquiver}
U(k_1)_{N_1-k_2}\times U(k_2)_{N_2-k_2} - \text{BF}\quad\Leftrightarrow\quad U(N_1-k_2)_{-k_1} \times U(N_2-k_1)_{-k_2} +\text{BF}.
\end{align}
This is obtained from gauging the diagonal $U(1)$ symmetry of two copies of the level-rank duality \eqref{eq:lr 2r}, which is described in more detail in Appendix \ref{app:two-node}.

\section{Adjoint Dualities via Orbifolding}
\label{sec:orbs}

We now consider the two-node quiver we derived in Sec. \ref{subsec:Flavor-Violated 2r}, with $N_{1}=N_{2}=N$ and $k_{1}=k_{2}=k$ where $N>k$ due to the flavor bound. These parameters are more constrained for the flavor-bounded case where $k_{1}=N_{2}$ was required, which would imply $N=k$. What is special about these particular values is that the fermionic theories of this quiver have an explicit $\mathbb{Z}_{2}$ symmetry, which can be thought of as interchanging the two nodes. In this section we will argue this special class of quivers allows us to re-derive the rank-two bosonization duality with real adjoint matter on each side \cite{Gomis:2017ixy}. First, let us briefly review the adjoint matter dualities.

\subsection{Adjoint 3d Bosonization Duality}
\label{sec:adjoint_rev}

\begin{figure}
\begin{centering}
\includegraphics[scale=0.6]{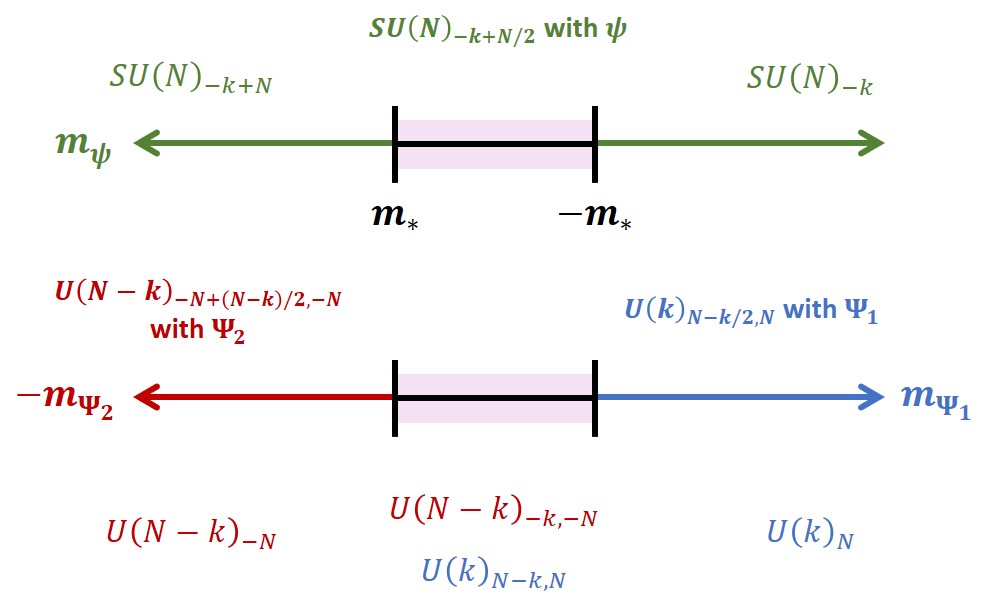}
\par\end{centering}
\caption{Phase diagram of the 3d bosonization duality with real adjoint matter \cite{Gomis:2017ixy}. Distinct theories are shown in different colors. The region shaded in purple corresponds to the quantum phase. \label{fig:Real-adjoint-phase 2r}}
\end{figure}

Ref. \cite{Gomis:2017ixy} conjectured the phases of QCD$_{3}$ adjoint.\footnote{Ref. \cite{Gomis:2017ixy} uses a slightly different notation than we use here. Their results can be recovered by $k\rightarrow k-N/2$ with the overall time-reversal which flips the signs of the Chern-Simons levels.} Schematically, the fermion side is
\begin{equation}
SU(N)_{-k+N/2}\text{ with adjoint \ensuremath{\psi}},
\end{equation}
with $N>k$. Mass deforming the adjoint fermion gives $SU(N)_{-k}$ and $SU(N)_{-k+N}$. When $N\leq k$, these two TFTs are conjectured to have no quantum region in between them. However, when $N>k$ an intermediate Grassmannian phase is conjectured to exist between the asymptotic semi-classical regions. The claim is that once again two different dual theories describe the fixed points at the edge of the Grassmannian. These $U$ side theories are
\begin{equation}
SU(N)_{-k+N/2}\text{ with \ensuremath{\psi^\text{adj}}}\qquad\leftrightarrow\qquad\begin{cases}
U(k)_{N-\frac{k}{2},N}\text{ with  \ensuremath{\Psi_1^\text{adj}}} & m_{\psi^\text{adj}}=-m_{*}\\
U(N-k)_{-\frac{N}{2}-\frac{k}{2},-N}\text{ with \ensuremath{\Psi_2^\text{adj}}} & m_{\psi^\text{adj}}=m_{*}.
\end{cases}\label{eq:adjoint 2r}
\end{equation}
This is qualitatively similar to the flavor-violated bosonization duality and is summarized in Fig. \ref{fig:Real-adjoint-phase 2r}. For the quantum phase, we find the TFTs $U(k)_{N-k,N}$ and $U(N-k)_{-k,-N}$, which as expected are level-rank dual by \eqref{eq:lr 2r u u}. Note the quantum phases on the $SU$ side has no good description in terms of the $SU$ theory -- it is best described by the $U$ side of the theory. This is also a feature of the flavor-violated duality \eqref{eq:aharony fv} and we will see similar features below.

%One thing to note about the $U$ side of this duality is that the adjoint matter does not couple to the $U(1)$ subgroup of $U(k)$. As such, qualitatively the $SU$ side is identical to the $U$ side, up to the additional decoupled $U(1)$ Chern-Simons term.

\subsection{Adjoint Duality via Orbifolding} \label{subsec:adjointorbifold}

As mentioned in the introduction, what is special about the $N_1=N_2=N$ and $k_1=k_2=k$ subclass of two-node quivers is that they present a possible means of constructing the rank-two matter dualities. More explicitly, since both sides of the dualities exhibit ``theory space'' $\mathbb{Z}_{2}$ symmetries, we can orbifold said symmetries on both sides of the duality to obtain a new conjectured duality. In this section, we show the rank-two dualities involving adjoint matter can be obtained by orbifolding\footnote{Orbifolding is a technique which projects a theory with a certain symmetry onto an invariant subspace. It is perhaps most familiar in the string theory literature, but such techniques have also been adapted to quantum field theory \cite{Kachru:1998ys, Lawrence:1998ja} in the context of planar equivalence between mother and daughter theories \cite{Bershadsky:1998cb,Kovtun:2004bz} or in the context of constructing manifestly supersymmetric lattice gauge theories \cite{Cohen:2003xe}. In a quantum field theory orbifolding is simply a procedure to produce a new theory (the ``daughter'') from an old one (the ``mother'') by projecting out all degrees of freedom not invariant under a specified discrete group. If two theories flow to one and the same IR fixed point and this fixed point possesses a $\mathbb{Z}_2$ symmetry, then obviously after orbifolding we once again obtain a unique IR theory and so we inherit a duality of the daughters from the duality of the mothers. The only way this could fail is if the $\mathbb{Z}_2$ symmetry of the IR is not manifest in the UV theories or different from the apparent UV symmetries.} different two-node quiver theories.

\begin{figure}
\begin{centering}
\includegraphics[scale=0.6]{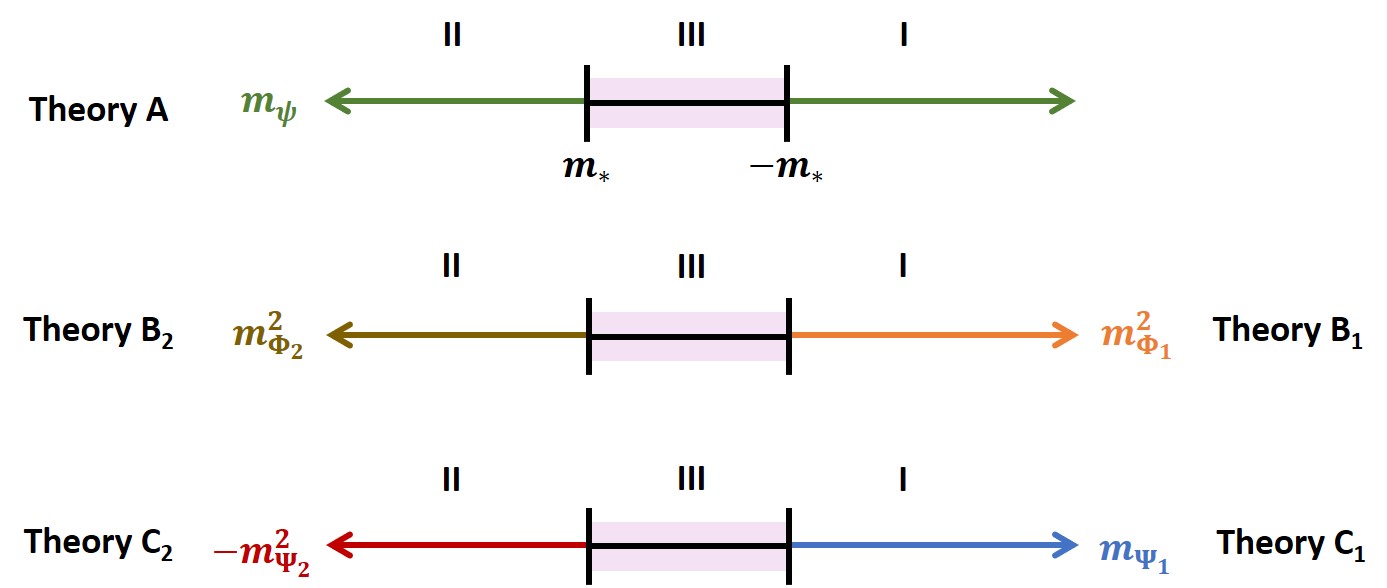}
\par\end{centering}
\caption{Phase diagram of the two-node quiver with explicit $\mathbb{Z}_{2}$ symmetry. Distinct theories are shown in different colors. Note the splitting of the bottom two phase diagrams which results from the use of the flavor-violated 3d bosonization duality. \label{fig:z2 sym 2r}}
\end{figure}

Fig. \ref{fig:z2 sym 2r} summarizes the full phase diagram of the $\mathbb{Z}_2$ symmetric quiver, including the intermediate scalar theories. The phases of Theory A are given by
\begin{subequations}
\begin{align}
\text{(I)}:\qquad & SU(N)_{-k} \times SU(N)_{-k} \\
\text{(II)}:\qquad & SU(N)_{-k+N} \times SU(N)_{-k+N} \\
\text{(III)}:\qquad &\text{Better described by $U$ side.}
\end{align}
\end{subequations}
Meanwhile, theories C$_1$ and C$_2$ describe the $U$ side, given by
\begin{subequations}
\begin{align}
\text{(I)}:\qquad & U(k)_{N} \times U(k)_{N} \\
\text{(II)}:\qquad & U(N-k)_{-N} \times U(N-k)_{-N} \\
\text{(III C$_1$)}:\qquad & U(k)_{N-k} \times U(k)_{N-k}+\text{BF} \\
\text{(III C$_2$)}:\qquad & U(N-k)_{-k}\times U(N-k)_{-k}-\text{BF}.
\end{align}
\end{subequations}
Similar to what we saw in Sec. \ref{subsec:Flavor-Violated 2r}, the BF terms in the phase III make the phases (III C$_1$) and (III C$_2$) level/rank dual to one another, as shown in Appendix \ref{app:two-node}. Observe that these phases already resemble ``two copies'' of the phases of the adjoint duality, up to certain subtleties in the qunatum phase, see Fig. \ref{fig:Real-adjoint-phase 2r}. Schematically, orbifolding causes the bifundamental fermions to become Majorana adjoint fermions and takes the TFTs from $G_P \times G_P \to G_P$, where $G_P$ is a gauge group $G$ with Chern-Simons level $P$, up to subtleties with Abelian levels. Orbifolding a $\mathbb{Z}_2$ symmetry to change a bifundamental to an adjoint representation has been previously explored in Ref. \cite{Kovtun:2007py, Unsal:unpub}.

More explicitly, the $\mathbb{Z}_{2}$ symmetric theory describing the $m_{\psi}<0$ half of the phase diagram is given by
\begin{subequations}
\label{eq:Z_2 sym lag}
\begin{align}
\mathcal{L}_{\text{A}} & =i\bar{\psi}\slashed{D}_{b_{1}-b_{2}+\tilde{A}_{1}}\psi-i\left[\frac{N-k}{4\pi}\text{Tr}_{N}\left(b_{1}db_{1}-i\frac{2}{3}b_{1}^{3}\right)\right] \nonumber
\label{eq:Z_2 su side} \\
& -i\left[\frac{N-k}{4\pi}\text{Tr}_{N}\left(b_{2}db_{2}-i\frac{2}{3}b_{2}^{3}\right)+\frac{N^{2}}{4\pi}\tilde{A}_{1}d\tilde{A}_{1}\right],\\
\mathcal{L}_{\text{C}_1} & =i\bar{\Psi}_1\slashed{D}_{c_{1}-c_{2}+\tilde{A}_{1}}\Psi_1-i\left[\frac{N}{4\pi}\text{Tr}_{k}\left(c_{1}dc_{1}-i\frac{2}{3}c_{1}^{3}\right)+\frac{N}{4\pi}\text{Tr}_{k}\left(c_{2}dc_{2}-i\frac{2}{3}c_{2}^{3}\right)\right],\label{eq:Z_2 u side}
\end{align}
\end{subequations}
where we have relabeled the gauge fields belonging to the first and second node of the $SU$ side as $b_{\mu}^{1}$ and $b_{\mu}^{2}$, respectively. On the $U$ side, $c_{\mu}^{1}$ and $c_{\mu}^{2}$ are the gauge fields of the first and second nodes. $\tilde{A}_1$ is the background gauge field associated with the $U(1)$ global symmetry.

The bifundamental fermion is in the representation $(\mathbf{N},\mathbf{\bar{N}})$. The $\mathbb{Z}_{2}$ symmetry acts as
\begin{equation}
\label{eq:su z2}
\mathbb{Z}_{2}:\qquad\psi\to-\psi^{\dagger},\qquad b_{\mu}^{1}\to b_{\mu}^{2},\qquad b_{\mu}^{2}\to b_{\mu}^{1}.
\end{equation}
Similarly, on the $U$ side, the theory is invariant under the symmetry transformation
\begin{equation}
\label{eq:u z2}
\mathbb{Z}_{2}:\qquad\Psi_a\to-\Psi_a^{\dagger},\qquad c_{\mu}^{1}\to c_{\mu}^{2},\qquad c_{\mu}^{2}\to c_{\mu}^{1}
\end{equation}
for $a=1,2$ corresponding to Theories C$_1$ and C$_2$. If we orbifold with respect to these $\mathbb{Z}_{2}$ symmetries on either side of the duality, the resulting daughter theories should continue to be dual.

We can arrive at the new Lagrangians describing the daughter theories by projecting the terms in \eqref{eq:Z_2 sym lag} to those invariant under \eqref{eq:su z2} and \eqref{eq:u z2}. On the $SU$ side, this effectively ties the $b_1$ and $b_2$ fields together, meaning the bifundamental now transforms as an adjoint, as desired. Note that this orbifold also breaks the $U(1)$ global symmetry associated with the phase of the matter to $\mathbb{Z}_{2}$, i.e. $\psi\to-\psi$.

The behavior of the Chern-Simons terms under the orbifold is slightly subtle. Since the action of orbifold projections on matter is well-known, it will be helpful to view the Chern-Simons term as arising from integrating out heavy fermions. This is directly analogous to the ``fiducial fermion'' presecription used to analyze the 3d bosonization dualities in the presence of a boundary \cite{Aitken:2017nfd, Aitken:2018joi}. For example, the $U(k)_{N}$ Chern-Simons term of a given node can arise from $N$ ``fiducial'' fermions with a large negative mass. In this case, the Chern-Simons term in the Lagrangian is replaced with
\begin{align}
-i\left[\frac{N}{4\pi} \text{Tr}_k \left(cdc-i\frac{2}{3}c^3\right) \right]\qquad \Leftrightarrow \qquad
\lim_{|m_\chi|\to \infty} i\bar\chi^M \slashed{D}_c \chi_M + m_{\chi}\bar\chi^M \chi_M
\end{align}
where $\chi_M$ are the fiducial fermions with $M = 1,\ldots,N$.\footnote{For brevity, we have neglected the Pauli-Villars regulators which also accompany fiducial fermions. See Refs. \cite{Aitken:2017nfd, Aitken:2018joi} for more details regarding this construction.}

Adopting this prescription for the Lagrangians in \eqref{eq:Z_2 sym lag}, each of the two nodes on the $SU$ ($U$) side picks up an extra $N-k$ ($N$) fundamental-represention fermions, which we will collectively denote $\chi_i$ for $i=1,2$. The $\mathbb{Z}_2$ symmetry transformations of \eqref{eq:su z2} and \eqref{eq:u z2} must then be supplemented with $\chi_1 \to \chi_2$ and $\chi_2 \to \chi_1$. The orbifold projection reduces these additional matter fields to a single group of fundamental fermions under the invariant gauge group. If we then restore the Chern-Simons term one gets from integrating out the fiducial fermions, we see that we simply lose one of the Chern-Simons terms under our orbifold, i.e. $U(k)_{N} \times U(k)_{N} \to U(k)_{N}$. Note this is fairly different than naively setting $c_1 = c_2$ in \eqref{eq:Z_2 u side}, in which case one would arrive at a $U(k)_{2N}$ Chern-Simons term.

Another important subtlety is one needs to orbifold the Pauli-Villars regulators as well. Hence, what was effectively a bifundamental Pauli-Villars regulator under $SU(N)\times SU(N)$ becomes a real adjoint PV regulator under $SU(N)$. All this means is that the $\eta$-invariant terms obey a similar relation to the Chern-Simons levels discussed above.

Thus the orbifolding reduces the ``two copies'' of the Chern-Simons terms to a Chern-Simons matter theory with a single dynamical gauge field resembling the adjoint duality. However, in the adjoint bosonization duality the TFTs found in the quantum region have the non-Abelian levels shifted relative to the Abelian levels, see Fig. \ref{fig:Real-adjoint-phase 2r}. For example, for the $m_{\psi^\text{adj}}<0$ side, as we mass deform $\Psi_1^\text{adj}$ we move from $U(k)_N$ to $U(k)_{N-k, N}$. This is contrast to the $\mathbb{Z}_{2}$ symmetric quiver, where we see no such effect because fundamental representation fermions shift Abelian and non-Abelian levels on equal footing. To see how such a change can arise from orbifolding, consider the bifundamental fermion on the $U$ side. One can clearly see that the bifundamental fermions remain invariant under the subgroup where $\text{Tr}\left(c_{1}\right)=\text{Tr}\left(c_{2}\right)$. Thus, since we a projecting to precisely this subgroup via our orbifold, the remaining $U(1)$ part should decouple.

To see this at the Lagrangian level, note the coupling of said fermion is giving by $i\bar{\Psi}\slashed{D}_{-c+g+\tilde{A}_{1}}\Psi$, and so for negative mass deformation we get the Chern-Simons term
\begin{align}
i\mathcal{L}_{\text{shift}} & =\frac{-1}{4\pi}\text{Tr}_{k^{2}}\left[\left(c_{1}\mathds{1}_{k}-\mathds{1}_{k}c_{2}+\mathds{1}_{k^{2}}\tilde{A}_{1}\right)d\left(c_{1}\mathds{1}_{k}-\mathds{1}_{k}c_{2}+\mathds{1}_{k^{2}}\tilde{A}_{1}\right)+\ldots\right]\nonumber \\
 & = -\frac{k^2}{4\pi}c_{1}dc_{1}-\frac{k^2}{4\pi}c_{2}dc_{2}+\frac{1}{2\pi}\text{Tr}_{k}\left(c_{1}\right)d\text{Tr}_{k}\left(c_{2}\right) + \ldots
\end{align}
where ``$\ldots$'' represents additional terms not relevant for the cancellation. Absolutely vital to the story is the additional BF term we get from the cross term of the two dynamical $U(k)$ gauge fields. Under the orbifolding procedure, we have $c_{1}=c_{2}$. Notice that when this is enforced the BF term precisely cancels the Abelian Chern-Simons terms. The net result is that the Abelian part of the Chern-Simons level gets no overall shift. The same effect occurs in the PV regulator. This means the quantum phase description
\begin{equation}
U(k)_{N-k}\times U(k)_{N-k}\quad\xrightarrow{\text{orbifold}}\quad  U(k)_{N-k,N}
\end{equation}
This exactly matches the quantum phase of an adjoint fermion described in Sec. \ref{sec:adjoint_rev}.

Putting this altogether, we arrive at the orbifolded Lagrangian description
\begin{subequations}
\begin{align}
\mathcal{L}_{\psi} & =i\bar{\psi}\slashed{D}_{b}\psi-i\left[\frac{N-k}{4\pi}\text{Tr}_{N}\left(bdb-i\frac{2}{3}b_{1}^{3}\right)\right]\\
\mathcal{L}_{\Psi} & =i\bar{\Psi}\slashed{D}_{c}\Psi-i\left[\frac{N}{4\pi}\text{Tr}_{k}\left(cdc-i\frac{2}{3}c^{3}\right)\right]
\end{align}
\end{subequations}
where we have defined the $\mathbb{Z}_{2}$ invariant subgroup as $b_{1}=b_{2}\equiv b$ and $c_{1}=c_{2}\equiv c$. We have dropped the $\tilde{A}_{1}$ background field, because as mentioned above, the orbifolding breaks the $U(1)$ global symmetry down to $\mathbb{Z}_{2}$.\footnote{Because there is a decoupled $U(1)$ Chern-Simons theory, it may look like there still should be a $U(1)$ global symmetry associated with its monopole current. However, this is not the case because there is no matter charged under said $U(1)$ Chern-Simons theory. Said another way, it is not possible to create gauge-invariant monopoles because there is no matter to cancel the GNO charge. }

\subsection{Novel Adjoint Dualities}

\begin{figure}
\begin{centering}
\includegraphics[scale=0.6]{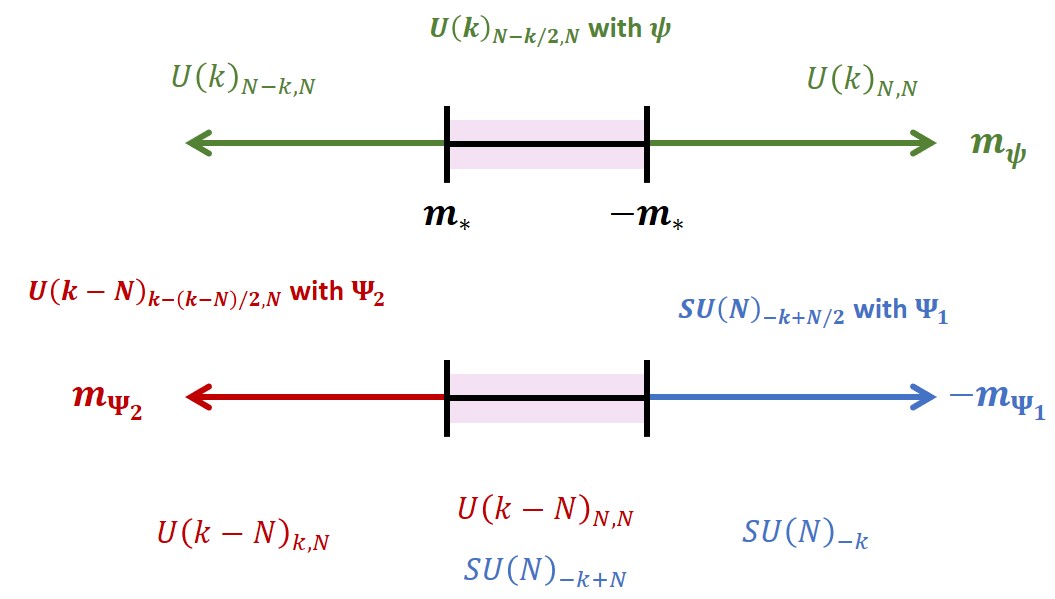}
\par\end{centering}
\caption{Phase diagram for the novel real adjoint bosonization duality.\label{fig:novel adjoint 2r}}
\end{figure}

We saw above that we can recover the rank-two duality of Ref. \cite{Gomis:2017ixy} by considering the two-node fermionic quiver when one of the 3d bosonization dualities is in the flavor-violated regime. In particular, in Sec. \ref{subsec:Flavor-Violated 2r} we considered the case where the Theory A to B duality was flavor-violated, while the Theory B to C duality was flavor-bounded, which brought us into the regime of $k<N<N_{*}$. A natural next step is to consider the opposite case, where now the Theory B to C duality is flavor-violated and the Theory A to B duality is flavor-bounded. This corresponds to the regime $N<k<N_{*}$ and gives a new $\mathbb{Z}_{2}$ symmetric quiver.

Working though the same construction we performed in Sec. \ref{subsec:Flavor-Violated 2r}, but instead beginning on the $U$-$U$ side, we can construct a general flavor-violated two-node quiver and then specialize to the $\mathbb{Z}_{2}$ symmetric case. The fact that we are starting with a $U$-$U$ theory on one end complicates things slightly because of the presence of mixed BF terms coming from the $\eta$-invariants in the mass deformed phases. The A, $\text{C}_{1}$, and $\text{C}_{2}$ theory again exhibit a manifest $\mathbb{Z}_{2}$ symmetry. If one again orbifolds with respect to said $\mathbb{Z}_{2}$ symmetry, we arrive at a new adjoint duality, given by
\begin{equation}
U(k)_{N-k/2,N}\text{ with \ensuremath{\psi^\text{adj}}}\qquad\leftrightarrow\qquad\begin{cases}
SU(N)_{-k+N/2}\text{ with  \ensuremath{\Psi_1^\text{adj}}} & m_{\psi^\text{adj}}=m_{*}\\
U(k-N)_{k-\left(k-N\right)/2,N}\text{ with \ensuremath{\Psi_2^\text{adj}}} & m_{\psi^\text{adj}}=-m_{*}
\end{cases}\label{eq:new adjoint 2r}
\end{equation}
where now we require $k>N$. The phase diagram for said duality is shown in Fig. \ref{fig:novel adjoint 2r}. Once more, the $\mathbb{Z}_{2}$  projection causes some of the BF terms to cancel $U(1)$ Chern-Simons terms, making the $U(1)$ Chern-Simons theories decoupled from the adjoint matter.

What is unique about this adjoint duality is that one side of the phase diagram is described by both an $SU$ and $U$ theory. One can see this is necessary from the start. Mass deforming the original $U$ theory gives the TFTs $U(k)_{N,N}$ and $U(k)_{N-k,N}$. Via the level-rank dualities, \eqref{eq:lr 2r} and \eqref{eq:lr 2r u u}, the first of these is dual to an $SU$ theory while the second is dual to a $U$ theory. Despite this difference, the two theories agree in the quantum region, since the $U$ theory yields a TFT which is level-rank dual to an $SU$ TFT.

There is an alternative way to arrive at this duality which also lends insight into the relationship between \eqref{eq:adjoint 2r} and \eqref{eq:new adjoint 2r}. The $\mathbb{Z}_{2}$-symmetric quiver of Fig. \ref{fig:z2 sym 2r} has a $U(1)$ global symmetry at every step, which we call $\tilde{A}_{1}$. If we gauge said symmetry, we would pick up a new $U(1)$ global monopole symmetry which couples to $\tilde{A}_{1}$ through a BF term, which we call $\tilde{B}_{1}$. If we \emph{also }gauge this $U(1)$ symmetry, we claim we arrive at the very same $\mathbb{Z}_{2}$-symmetric quiver which produced Fig. \ref{fig:novel adjoint 2r}, up to a time-reversal and label interchange $N\leftrightarrow k$.

How this occurs is straightforward to see qualitatively. On the $SU(N)_{-k+N/2}\times SU(N)_{-k+N/2}$ side of the duality, the two additional $U(1)$ gauge symmetries allow us to rewrite the theory as $U(N)_{-k+N/2}\times U(N)_{-k+N/2}$. Similarly, for the $U(k)_{N}\times U(k)_{N}$ theory, integrating out the two additional $U(1)$ gauge symmetries cancels the existing $U(1)\subset U(k)$ gauge groups, giving an $SU(k)_{N}\times SU(k)_{N}$ theory. Theory $\text{C}_{2}$ follows in a similar manner.\footnote{Here, we remark that the adjoint duality with unitary gauge group \eqref{eq:new adjoint 2r} can also be achieved by consistent $S$ and $T$ operations along the phase diagram of $SU(N)_k+\text{adjoint}$ theory in Ref. \cite{Gomis:2017ixy}. Although there is no $U(1)$ global symmetry acting faithfully, we could still introduce the $U(1)$ gauge field and couple it only to the gauge fields. Consistency of the result supports that the mapping of the $\mathbb Z_2$ symmetries along the orbifolding quiver is correct.}

Having considered $\mathbb{Z}_2$-symmetric quivers in the $N<k$ and $k<N$ regimes, the very last case we can consider consists of taking $N=k$. This corresponds to using two flavor-bounded dualities to derive the two-node quiver, and thus brings us back into the regime of the flavor-bounded duality of Sec. \ref{sec:flavor-bounded}. Again, we can $\mathbb{Z}_{2}$-orbifold, and we find
\begin{equation} \label{eq:newadj}
SU(N)_{-N/2}\text{ with \ensuremath{\psi^\text{adj}}}\qquad\leftrightarrow\qquad U(N)_{N/2,N}\text{ with  \ensuremath{\Psi^\text{adj}}}.
\end{equation}
Note there is no quantum phase for these theories, which is expected because no flavor-violated duality was needed in the construction of the relevant $\mathbb{Z}_2$ symmetric quiver.

\subsection{Dualities with Extra Flavors}

\begin{figure}
\begin{centering}
\includegraphics[scale=0.6]{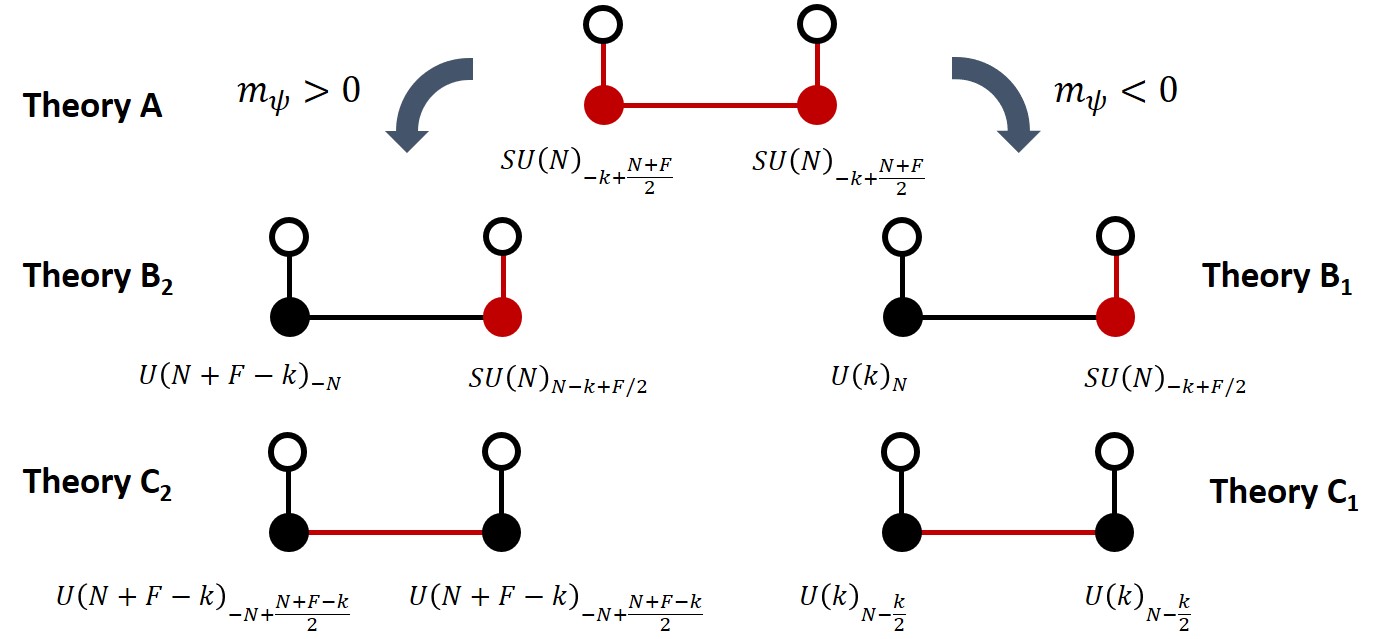}
\par\end{centering}
\caption{Phase diagram of the two-node quiver with extra flavors on each node and explicit $\mathbb{Z}_{2}$ symmetry. Here the white nodes represent \emph{global} flavor symmetries of the extra matter. Once more, the splitting of the bottom two phase diagrams results from the use of the flavor-violated 3d bosonization duality. We assume $N>k$ and $F<k$. \label{fig:z2 flavors 2r}}
\end{figure}

We might try to use this procedure to see if we can derive qualitatively different dualities. Because we required $\mathbb{Z}_{2}$ symmetry we have considered all relative values of $N$ and $k$, we need to look at new quiver configurations. Following Ref. \cite{Aitken:2019mtq}, a natural generalization is to add flavors on each of the two nodes. We do this in such a way to maintain the $\mathbb{Z}_{2}$ symmetry, but this could of course be done more generally. Starting with fermion flavors on each node, the quiver diagram for this procedure is summarized in Fig. \ref{fig:z2 flavors 2r}.

On the $SU$ side, we begin with the theory
\begin{equation}
SU(N)_{-k+(N+F)/2}\times SU(N)_{-k+(N+F)/2}+\text{bifund. }\psi+F\,\psi_{1}+F\,\psi_{2},
\end{equation}
where we have denoted by the extra flavor degrees of freedom belonging to node $i$ by $\psi_{i}$. This can then be dualized using a flavor-violated version of Aharony's duality. The application of said duality changes one of the nodes to a $U$ group and both the bifundamental and corresponding flavor degree of freedom into scalars.

Since the second node has both fermionic and bosonic degrees of freedom, one can make use of the master duality to dualize.\footnote{See Refs. \cite{Aitken:2018cvh, Aitken:2019mtq} for more details on how to dualize quivers using the master duality.} Before doing so, a subtle point is that having both scalars and fermions charged under the same node naturally gives rise to scalar-fermion interactions. These are identical in form to the interaction of the master bosonization duality \cite{Jensen:2017bjo,Benini:2017aed}. Specifically, these are interactions of the form
\begin{equation}
\mathcal{L}_{\text{int}}\supset\left(\phi_{A\alpha_{2}}^{\dagger}\psi_{2}^{\alpha_{1}\alpha_{2}}\right)\left(\bar{\psi}_{\alpha_{1}\beta_{2}}^{2}\phi^{A\beta_{2}}\right).
\end{equation}
with $\alpha_{1},\beta_{1}$ indices of the $SU(N)$ gauge symmetry associated with the first node, $\alpha_2,\beta_2$ indices of the second node, and $A,B$ flavor indices.

\begin{figure}
\begin{centering}
\includegraphics[scale=0.58]{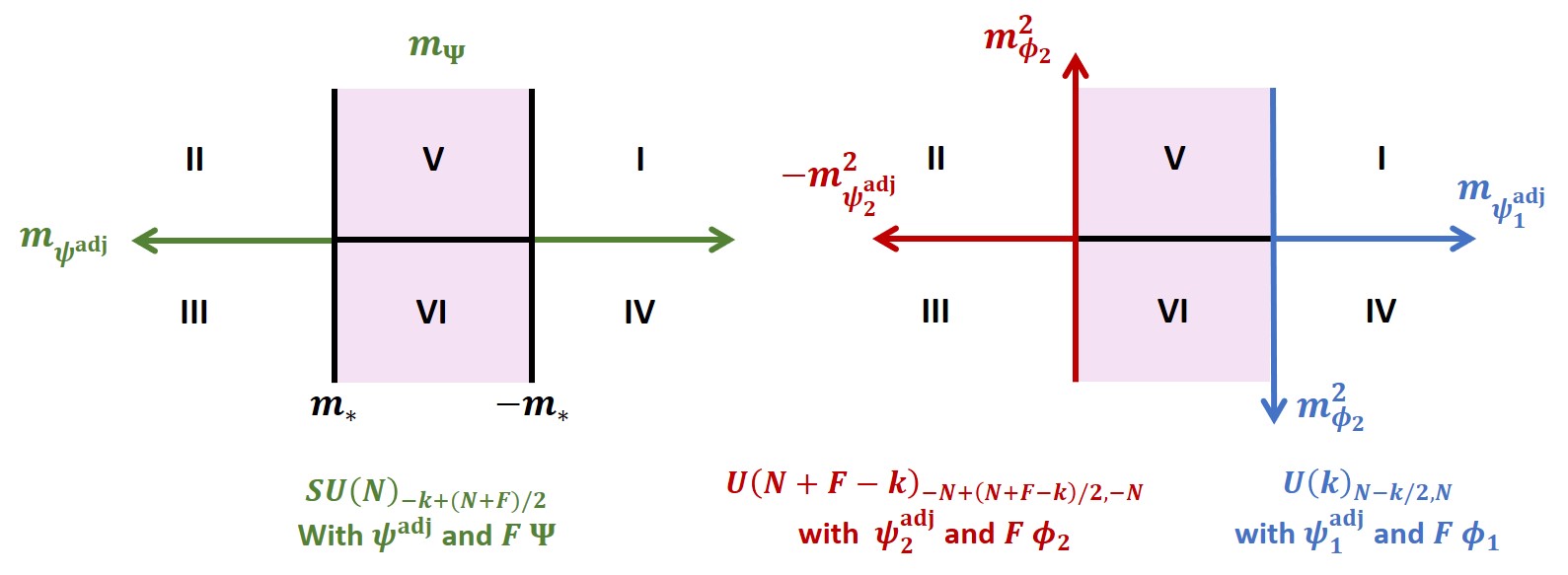}
\par\end{centering}
\caption{Phase diagram of the 3d bosonization duality with both adjoint and fundamental representation matter. Here we assume $N>k$ and $F<k$. \label{fig:fund and adj 2r}}
\end{figure}

Dualizing the second node also turns it into a $U$ group, and changes the scalars (fermions) degrees of freedom to fermions (scalars). Hence, we ultimately end up with a two-node quiver coupled via a bifundamental fermion with each of the nodes having scalar flavor degrees of freedom. Once more, the scalar and fermion degrees of freedom of each node have interactions of the form
\begin{equation}
\mathcal{L}_{\text{int}}\supset\left(\bar{\psi}_{\alpha_{1}\alpha_{2}}\Phi_{1}^{\alpha_{1}A}\right)\left(\Phi_{\beta_{1}A}^{1\dagger}\psi^{\beta_{1}\alpha_{2}}\right)+\left(\bar{\psi}_{\alpha_{1}\alpha_{2}}\Phi_{2}^{A\alpha_{2}}\right)\left(\Phi_{A\beta_{2}}^{2\dagger}\psi^{\alpha_{1}\beta_{2}}\right)
\end{equation}
with the same convention for coefficients used above. In order for the asymptotic phases to match the interaction must come with a positive coefficient. The net effect of this term is that when the scalars acquire a vacuum expectation value, a subgroup of the bifundamental fermion acquires a positive mass. In particular, with the assumption of maximal Higgsing, we have a breaking of $U(k)\to U(k-F)\times U(F)$ (or $U(N+F-k)\to U(N-k)\times U(F)$ on the $m_{\psi}>0$ side) . The mass term is of the form
\begin{equation}
\mathcal{L}_{\text{int}}\supset\left(\begin{array}{cc}
\mathds{1}_{F}\\
 & 0
\end{array}\right)_{\beta_{1}}^{\alpha_{1}}\bar{\psi}_{\alpha_{1}\alpha_{2}}\psi^{\beta_{1}\alpha_{2}}+\left(\begin{array}{cc}
\mathds{1}_{F}\\
 & 0
\end{array}\right)_{\beta_{2}}^{\alpha_{2}}\bar{\psi}_{\alpha_{1}\alpha_{2}}\psi^{\alpha_{1}\beta_{2}}
\end{equation}
so that it is the $U(F)$ subgroup which acquires a mass. Note from the Higgsing pattern above, these fermions are no longer charged under the dynamical symmetry and thus are sometimes referred to as ``singlets''. Such interactions are necessary to get a matching of the phase diagrams across the duality, see Fig. \ref{fig:fund and adj 2r}.

We now orbifold with respect to the very same $\mathbb{Z}_{2}$ symmetry we used above of the adjoint case, with the additional identifications $\psi_1 \Leftrightarrow \psi_2$ and $\Phi_1 \Leftrightarrow \Phi_2$. This is analogous to the orbifolding of the fiducial fermions used earlier.  We arrive at a duality which has both adjoint and fundamental representation matter on each side,
\begin{multline}
SU\left(N\right)_{-k+\left(N+F\right)/2}\text{ with \ensuremath{\psi^\text{adj}} and \ensuremath{F} \ensuremath{\psi^{\prime}}}\qquad\leftrightarrow\qquad\\
\begin{cases}
U(k)_{N-k/2,N}\text{ with \ensuremath{\Psi_1^\text{adj}}and \ensuremath{F} \ensuremath{\phi_{1}}} & m_{\psi^\text{adj}}=-m_{*},\\
U(N+F-k)_{-N+\left(N+F-k\right)/2,-N}\text{ with \ensuremath{\Psi_2^\text{adj}}and \ensuremath{F} \ensuremath{\phi_{2}}} & m_{\psi^\text{adj}}=m_{*}.
\end{cases}
\end{multline}
Note the interactions between the adjoint matter and the scalars is still present, e.g. $\left(\bar{\Psi}_{1\beta}^{\alpha}\phi_{1}^{\gamma A}\right)\times\left(\phi_{1\alpha A}^{\dagger}\Psi_{1\gamma}^{\beta}\right)$, with $\alpha, \beta, \gamma$ the $U(k)$ gauge indices and $A$ the $U(F)$ flavor index.

The phase diagram for this duality is shown in Figure \ref{fig:fund and adj 2r}. The phases on the $SU$ side are given by
\begin{subequations}
\begin{align}
\text{(I)}:\qquad & SU(N)_{-k+F} \\
\text{(II)}:\qquad & SU(N)_{-k+F+N} \\
\text{(III)}:\qquad & SU(N)_{-k+N} \\
\text{(IV)}:\qquad & SU(N)_{-k} \\
\text{(V), (VI)}:\qquad &\text{Better described by $U$ side.}
\end{align}
\end{subequations}
Meanwhile, the theories on the $U$ side are given by
\begin{subequations}
\begin{align}
\text{(I)}:\qquad & U(k-F)_{N} \\
\text{(II)}:\qquad & U(N+F-k)_{-N} \\
\text{(III)}:\qquad & U(N-k)_{-N} \\
\text{(IV)}:\qquad & U(k)_{N} \\
\text{(V)}:\qquad & U(k-F)_{N-k+F,N}\quad\Leftrightarrow\quad U(N+F-k)_{F-k,-N} \\
\text{(VI)}:\qquad & U(k)_{N-k,N} \quad\Leftrightarrow\quad U(N-k)_{-k,-N}.
\end{align}
\end{subequations}

\section{Discussion and Conclusion}
\label{sec:conclusion}

In this work we showed that we can connect the rank-one matter bosonization dualities to many dualities which contain adjoint matter for the unitary group case. It turns out that generalization of the two-node quiver dualities to the orthogonal and symplectic group is straightforward using both flavor-bounded and flavor-violated bosonization duality in Refs. \cite{Aharony:2016jvv,Komargodski:2017keh}. We describe these results in the Appendix \ref{app:so/sp}. The dualities of adjoint fermions in the orthogonal/symplectic group cases are distinguished from the unitary group case since the matter representation of the dual theory is not adjoint but symmetric-traceless/antisymmetric-traceless, respectively. Thus the natural next question is whether or not we can connect them to the rank-two matter dualities which contain symmetric and antisymmetric matter.

\begin{figure}
\begin{centering}
\includegraphics[scale=0.6]{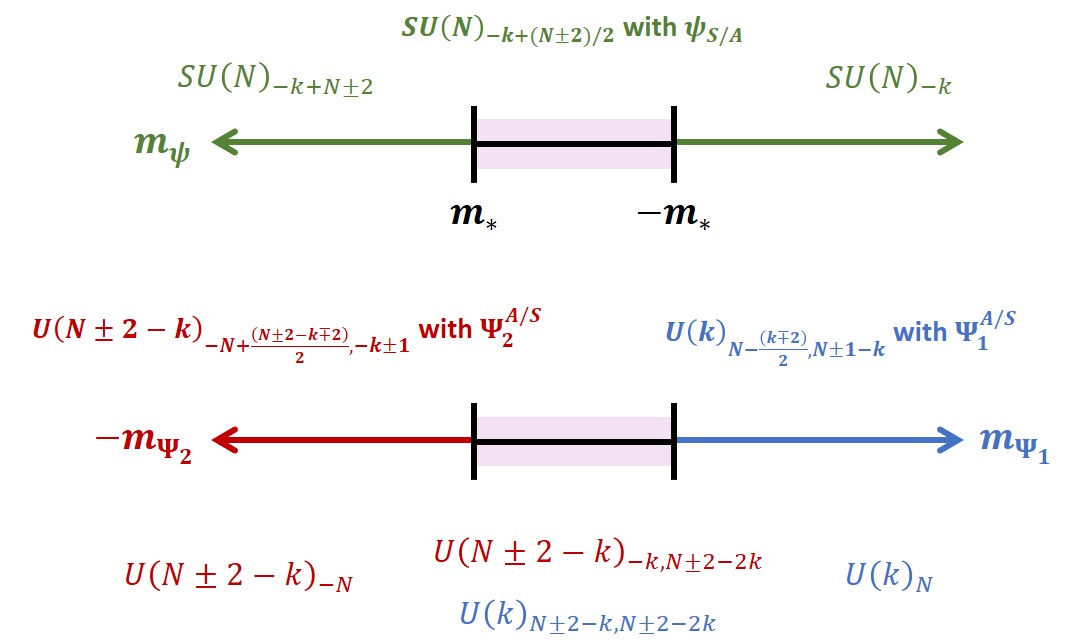}
\par\end{centering}
\caption{Symmetric and antisymmetric representation phase diagram.\label{fig:Symmetric-and-antisymmetric}}
\end{figure}

Ref. \cite{Choi:2018tuh} conjectured a similar duality between symmetric and antisymmetric matter for the unitary group case, the phase diagram of which is summarized in Fig. \ref{fig:Symmetric-and-antisymmetric}. Note the difference in the levels on the $SU$ is determined by (twice) the Dynkin index, $T(\text{S/AS})=\frac{1}{2}(N\pm2)$. In order for all the levels to match on the $U$ side, one must introduce the opposite type of matter, giving the dualities
\begin{equation}
SU(N)_{-k+\frac{1}{2}(N\pm2)}\text{ with \ensuremath{\psi^{\text{S/AS}}}}~\leftrightarrow~\begin{cases}
U(k)_{N-\frac{1}{2}(k\mp2),N-(k \mp 1)}\text{ with }\,\Psi^{\text{AS/S}}_{1} & m_{\psi}=-m_{*}\\
U(N\pm2-k)_{-N+\frac{1}{2}(N\pm2+k\mp2),\pm1-k}\text{ with }\,\Psi^{\text{AS/S}}_{2} & m_{\psi}=m_{*}
\end{cases}
\end{equation}
where the superscript S/AS denotes symmetric/antisymmetric representation under the gauge group.

The analog of orbifold producing daugther theory with symmetric or antisymmetric matter is called an ``orientifold''. For the bifundamental fermion in the $(\mathbf{N},\mathbf{N})$ representation, corresponding $\mathbb{Z}_{2}$ symmetries of the orientifold projection is given by\footnote{Note that even in the $(\mathbf{N},\mathbf{\bar{N}})$ case there exists another $\mathbb{\mathbb{Z}}_{2}$ symmetry
\begin{equation}
\mathbb{\mathbb{Z}}_{2}^{\prime\text{adj}}:\qquad\psi\to\psi^{\dagger},\qquad b_{\mu}^{1}\to b_{\mu}^{2},\qquad b_{\mu}^{2}\to b_{\mu}^{1}.
\end{equation}
In this respect both cases are on an equal footing. The difference here is the fact that in the $(\mathbf{N},\mathbf{\bar{N}})$ case it doesn't matter which $\mathbb{Z}_{2}$ we orbifold with respect to on either side; both lead to an adjoint rank-two tensor.} \cite{Unsal:unpub}
\begin{align}
\mathbb{Z}_{2}^{\text{S}}:\qquad & \psi\to\psi^{T},\qquad b_{\mu}^{1}\to b_{\mu}^{2},\qquad b_{\mu}^{2}\to b_{\mu}^{1},\\
\mathbb{Z}_{2}^{\text{AS}}:\qquad & \psi\to-\psi^{T},\qquad b_{\mu}^{1}\to b_{\mu}^{2},\qquad b_{\mu}^{2}\to b_{\mu}^{1}.
\end{align}
Constructing two-node quiver dualities with $(\mathbf{N},\mathbf{N})$ representation matter is straightforward. Much of the derivation of the $(\mathbf{N},\mathbf{\bar{N}})$ case is mirrored here, but now the matter will simply be fundamentally charged under both gauge groups at every step of the way.

The naive application of the above procedure is not successful. The difficulty lies in realizing $\mathcal{O}(1)$ shifts of the rank of the gauge groups which are required for the daughter dualities. We have not yet been able to find where the required finite shifts in the gauge group come from. Similar subtleties arise in the case of the $SO/Sp$ gauge theory with rank-two fermions. It would be interesting to analyze the orbifold/orientifold equivalence in the Chern-Simons matter theories up to the subleading $\mathcal{O}(1)$ order to resolve the issue. Another interesting question is why the unitary group rank-two dualities map representations as adjoint to adjoint and symmetric to antisymmetric.\footnote{The symplectic (orthogonal) dualities follow this latter pattern since in this case the adjoint is the symmetric (antisymmetric) to begin with.} It would be nice to understand the origin of this mapping carefully from the orbifolding point of view. We leave these puzzles for future work.

Despite the difficulty in connecting such dualities to the two-node quivers, we can follow a similar procedure as we did above for constructing new adjoint dualities to slightly generalize the symmetric/antisymmetric dualities. For instance, we can conjecture the new duality
\begin{multline}
U(k)_{N-\frac{1}{2}(k\pm2),N-k\mp1}\text{ with \ensuremath{\psi^{\text{S/AS}}}}\\~\leftrightarrow~
\begin{cases}
SU(N)_{-k+\frac{1}{2}(N\mp2)}\text{ with }\,\Psi_{1}^{\text{AS/S}} & m_{\psi}=-m_{*}\\
U(k\pm2-N)_{k-\frac{1}{2}(k\pm2-N\mp2),k\pm1}\text{ with }\,\Psi^{\text{AS/S}}_{2} & m_{\psi}=m_{*}
\end{cases}
\end{multline}
which must obey $k>N\pm2$. This passes many of the same consistency checks as the original symmetric/antisymmetric duality. Furthermore, limiting case $k=T(R)$ gives the new symmetric/antisymmetric duality with no quantum phase,
\begin{equation}
\begin{aligned}
SU(N)_{-(N\pm2)/2}\text{ with \ensuremath{\psi^\text{S/A}}}\qquad\leftrightarrow\qquad U(N\pm2)_{N/2,\mp 1}\text{ with  \ensuremath{\Psi^\text{A/S}}},
\end{aligned}
\end{equation}
which is qualitatively similar to the new adjoint duality found in \eqref{eq:newadj}.

%It is very tempting to say this flavored duality has some relation to the symmetric/antisymmetric quiver. Note that if we set $F=2$ and look at just the top-left and bottom-right phases of Fig \ref{fig:fund and adj 2r}, the phases are almost identical to what we would expect for the symmetric duality, see Fig. \ref{fig:Symmetric-and-antisymmetric 2r}. Unfortunately it does not appear the Grassmannian phases match. Maybe there is something more here?

\section*{Acknowledgments}

We would like to thank Mithat \"Unsal and Laurence Yaffe for sharing unbpublished work, \cite{Unsal:unpub}. The work of KA and AK was supported, in part, by the U.S.~Department of Energy under Grant No.~DE-SC0011637 and the Simons Foundation as part of the Simons Collaboration on Ultra Quantum Matter. CC is supported in part by the Simons Foundation grant 488657 (Simons Collaboration on the Non-Perturbative Bootstrap). Any opinions, findings, and conclusions or recommendations expressed in this material are those of the authors and do not necessarily reflect the views of the funding agencies.

\begin{appendix}

\section{Deriving Two-Node Fermionic Dualities}
\label{app:two-node}

In this appendix we give the details of the two-node quivers constructed in Sec. \ref{sec:two-node}.

We begin with the flavor-bounded case. In the main text, we showed a two-node fermionic quiver can be found by matching the scalar sides of Aharony's dualities \eqref{eq:aharony}, with appropriate relabeling and background terms. To achieve this matching, it is useful to work with the charge conjugated version of \eqref{eq:aharony 2}, which at the Lagrangian level is given by
\begin{subequations}
\label{eq:aharony lag 2}
\begin{align}
\mathcal{L}_{SU} & =\left|D_{-b^{\prime}+B+\tilde{A}_{1}}\phi\right|^{2}-i\left[-\frac{k}{4\pi}\text{Tr}_{N}\left(b^{\prime}db^{\prime}-i\frac{2}{3}b^{\prime3}\right)-\frac{Nk}{4\pi}\tilde{A}_{1}d\tilde{A}_{1}\right],\\
\mathcal{L}_{U} & =i\bar{\Psi}\slashed{D}_{-c+B}\Psi-i\left[\frac{N}{4\pi}\text{Tr}_{k}\left(cdc-i\frac{2}{3}c^{3}\right)+\frac{N}{2\pi}\text{Tr}_{k}\left(c\right)d\tilde{A}_{1}\right],\label{eq:aharony 1 lag 2r}
\end{align}
\end{subequations}
where $B\in su(N_f)$ and the duality is subject to the flavor bound $N\geq N_{f}$ and mass mapping $m_{\Psi}\leftrightarrow m_{\phi}^{2}$.

To perform the matching, we rearrange the $U$ sides of \eqref{eq:aharony lag 1} and \eqref{eq:aharony lag 2} by shifting the dynamical field to eliminate the BF terms and add appropriate background Chern-Simons terms. For \eqref{eq:aharony 2 lag 2r} we take $\tilde{c}\to\tilde{c}+\tilde{A}_{1}$, giving
\begin{equation}
\mathcal{L}_{U}=\left|D_{c-C+\tilde{A}_{1}}\Phi\right|^{2}-i\left[\frac{N}{4\pi}\text{Tr}_{k}\left(cdc-i\frac{2}{3}c^{3}\right)-\frac{Nk}{4\pi}\tilde{A}_{1}d\tilde{A}_{1}\right].
\end{equation}
Additionally, we add $-i\left[\frac{Nk}{4\pi}\tilde{A}_{1}d\tilde{A}_{1}\right]$ to both sides of \eqref{eq:aharony lag 1} to cancel the $\tilde{A}_1$ background Chern-Simons term. For \eqref{eq:aharony 1 lag 2r}, we take $\tilde{c}\to\tilde{c}-\tilde{A}_{1}$ and then add $-i\left[\frac{Nk}{4\pi}\tilde{A}_{1}d\tilde{A}_{1}\right]$ to both sides.

At this point one can promote the background flavor symmetries in both dualities so all the matter is bifundmentally charged. If one compares the scalar sides of the two dualities, an explicit matching can be achieved by identifying the fields
\begin{subequations}
\begin{align}
SU\text{ field:} & & C & \qquad\Leftrightarrow\qquad b^\prime, \\
U\text{ field:} & & c & \qquad\Leftrightarrow\qquad G \equiv B + \tilde{A}_1 \mathds{1}_{N_s}, \\
U(1)\text{ background field:} & & \tilde{A}_1 & \qquad\Leftrightarrow\qquad \tilde{B}_1.
\end{align}
\end{subequations}
Thus we also arrive at a duality between the respective fermion dualities, which is our desired flavor-bounded two-node quiver. The mass mapping is such that $m_\psi$ is identified with $-m_\Psi$.

A self-consistency check can be performed by making sure the two sides of the duality still match under mass deformations. We find
\begin{subequations}
\begin{align}
\left(\text{A}\right)\qquad m_{\psi}<0:\qquad & SU(N_1)_{-k_{1}}\times SU(N_2)_{-k_{2}} \label{eq:mass_def_1} \\
\left(\text{C}\right)\qquad m_{\Psi}>0:\qquad & U(k_1)_{N_{1}}\times U(k_2)_{N_{2}}\label{eq:mass_def_2} \\
\left(\text{A}\right)\qquad m_{\psi}>0:\qquad & SU(N_1)_{-k_{1}+N_{2}}\times SU(N_2)_{-k_{2}+N_{1}}=SU(N_2)_{-k_{2}+N_{1}} \label{eq:mass_def_3} \\
\left(\text{C}\right)\qquad m_{\Psi}<0:\qquad & U(k_1)_{N_{1}-k_{2}}\times U(k_2)_{N_{2}-k_{1}}=SU(N_2)_{-k_{2}+N_{1}} \label{eq:mass_def_4}
\end{align}
\end{subequations}
where we have used the fact $k_{1}=N_{2}$ is required by the flavor constraints. Clearly \eqref{eq:mass_def_1} and \eqref{eq:mass_def_2} are level-rank dual to one another. Eqs. \eqref{eq:mass_def_3} and \eqref{eq:mass_def_4} take slightly more work. Since $k_{1}=N_{2}$, each of these theories has one of the gauge field's Chern-Simons terms vanish. In \eqref{eq:mass_def_3}, all degrees of freedom of the $SU(N_1)_0$ theory are gapped out in the IR limit, and we can thus drop this factor. For \eqref{eq:mass_def_4}, one can integrate out the $U(1)$ subgroups of $U(k_1)$ and $U(k_2)$. When one does this, both $U(1)$ factors get eliminated and the theory reduces to $SU(N_2)_{N_1-k_2} \times SU(k_2)_0$, which matches \eqref{eq:mass_def_3} after dropping the second term.

If we want to work in the regime where $k_1\ne N_2$, we would need to use the flavor-violated 3d bosonization duality in one of these steps. In the main text, we choose to replace \eqref{eq:aharony lag 1} with its flavor-violated equivalent. This extends the flavor bounds to $k_1 < N_2 < N_*$ but means the full phase diagram of the $U$ side of the duality is described by two separate scalar theories, corresponding to the $m_\psi>0$ and $m_\psi<0$ halves of the $SU$ phase diagram (this is analogous to the layout in Fig. \ref{fig:Real-adjoint-phase 2r}).

Fortunately, the $m_\psi<0$ theory is identical at the Lagrangian level to the flavor-bounded duality, i.e. \eqref{eq:aharony 2 lag 2r}. Thus the very same matching that was performed above for the flavor-bounded case can be repeated for this end of the phase diagram. The $m_\psi>0$ side requires one to perform the matching once more. The explicit form the Lagrangian for Theories $\text{B}_{2}$ and $\text{C}_{2}$ are given by
\begin{subequations}
\label{eq:flavor violated lag}
\begin{align}
\mathcal{L}_{\text{B}_{2}} & =\left|D_{c-C-\tilde{A}_{1}}\Phi_{2}\right|^{2}-i\left[\frac{-N_{1}}{4\pi}\text{Tr}_{N_{2}-k_{1}}\left(cdc-i\frac{2}{3}c^{3}\right)\right]\nonumber \\
 & -i\left[\frac{N_{1}-k_{2}}{4\pi}\text{Tr}_{N_{2}}\left(CdC-i\frac{2}{3}C^{3}\right)+\frac{N_{1}N_{2}}{4\pi}\tilde{A}_{1}d\tilde{A}_{1}\right],\\
\mathcal{L}_{\text{C}_{2}} & =i\bar{\Psi}\slashed{D}_{-c+g+\tilde{A}_{1}}\Psi-i\left[-\frac{k_{1}}{4\pi}\text{Tr}_{N_{1}-k_{2}}\left(cdc-i\frac{2}{3}c^{3}\right)-\frac{k_{2}}{4\pi}\text{Tr}_{N_{2}-k_{1}}\left(gdg-i\frac{2}{3}g^{3}\right)\right]\nonumber \\
 & -i\left[\frac{1}{2\pi}\text{Tr}_{N_{1}-k_{2}}\left(c\right)d\text{Tr}_{N_{2}-k_{1}}\left(g\right)+\frac{N_{1}N_{2}}{4\pi}\tilde{A}_{1}d\tilde{A}_{1}+\frac{\left(N_{1}-k_{2}\right)\left(N_{2}-k_{1}\right)}{4\pi}\tilde{A}_{1}d\tilde{A}_{1}\right],
\end{align}
\end{subequations}
where we have defined the field $g\in U(N_{2}-k_{1})$.

In order for the flavor-violated quiver to match in the intermediate phases, we detail the generalized level/rank duality between the two node quiver with BF term pointed out in \eqref{eq:levelrankquiver}. The main point as illustrated in Sec. \ref{sec:two-node} is that the gauging of diagonal $U(1)$ symmetry of the two copies of level/rank duality $SU(N_1)_{k_1}\times SU(N_2)_{k_2}\leftrightarrow U(k_1)_{-N_1}\times U(k_2)_{-N_2}$.

The explicit Lagrangian with two $U(1)$ background terms is given by \cite{Hsin:2016blu},
\begin{equation}
\begin{aligned}
~&-i\left [\frac{k_1}{4\pi}\text{Tr}_{N_1}(ada-i\frac{2}{3}a^3) +\frac{1}{2\pi}ed\text{Tr}_{N_1} (a+B) + \frac{k_2}{4\pi}\text{Tr}_{N_2}(bdb-i\frac{2}{3}b^3) +\frac{1}{2\pi}fd\text{Tr}_{N_2}(b+C)\right]
\Leftrightarrow
\\&-i\left [-\frac{N_1}{4\pi}\text{Tr}_{k_1}(udu-i\frac{2}{3}u^3) +\frac{1}{2\pi}\text{Tr}_{k_1}(u)dB-\frac{N_2}{4\pi}\text{Tr}_{k_2}(vdv-i\frac{2}{3}v^3) +\frac{1}{2\pi}\text{Tr}_{k_2}(v)dC\right].
\end{aligned}
\end{equation}
Now if we add mixed counter-term $-i[-\frac{1}{2\pi}BdC]$ along the duality and gauge the diagonal $U(1)$ as $B=C\rightarrow c$, we get following duality after solving the equation of motion
\begin{equation}
\begin{aligned}
~&-i\left [\frac{k_1}{4\pi}\text{Tr}_{N_1}(ada-i\frac{2}{3}a^3)+\frac{k_2}{4\pi}\text{Tr}_{N_2}(bdb-i\frac{2}{3}b^3)-\frac{1}{2\pi}\text{Tr}_{N_1}(a)d\text{Tr}_{N_2}(b)\right]
\\&\Leftrightarrow -i\left [ -\frac{N_1}{4\pi}\text{Tr}_{k_1}(udu-i\frac{2}{3}u^3)-\frac{N_2}{4\pi}\text{Tr}_{k_2}(vdv-i\frac{2}{3}v^3)+\frac{1}{2\pi}\text{Tr}_{k_1}(u)d\text{Tr}_{k_2}(v) \right].
\end{aligned}
\end{equation}
After change of variables, the above becomes equivalent to \eqref{eq:levelrankquiver}.

\subsection{Consistency Checks}

Another non-trivial consistency check of the quantum phase diagram of the flavor-violated two-node quiver can be established using similar approaches to the one in the literature, e.g. Refs. \cite{Hsin:2016blu,Cordova:2017vab,Benini:2017aed,Choi:2018tuh}. A useful check is the consistency of the background counterterms along the phase diagram, which we discuss in more detail in Appendix \ref{app:gravitational} for the case of gravitational counterterms.

Here, we show the non-trivial matching of the quantum phase of $SU(2)_k\times SU(2)_k+\psi^{\text{bifund}}$ and its isomorphic expression $Spin(4)_k+2~\psi^{\text{vec}}$ analyzed in Ref. \cite{Cordova:2017vab}. We will use explicit superscripts in this section to avoid confusion between vector and bifundamental matter. Since both of these theories use different dual descriptions -- the former with bifundamental scalar or fermion and the latter a vector scalar -- the matching of the intermediate phase mutually supports the two-node construction we have laid out above.

First, note the quantum phase exists only when $k=0$. From the viewpoint of the orthogonal gauge group, $Spin(4)_k+2~\psi^{\text{vec}}$ at small fermion mass flows to a non-linear sigma model with target space $S^1$ \cite{Cordova:2017vab}. On the other hand, the two-node fermionic quiver predicts $SU(2)_0 \times SU(2)_0+\psi^{\text{bifund}}$ flows to
\begin{align}
U(1)_{1}\times U(1)_{1} - \text{BF}\quad\leftrightarrow \quad U(1)_{-1}\times U(1)_{-1} +\text{BF}.
\end{align}
In terms of a $K$-matrix description, the phase is described by the 2 by 2 matrix
\begin{align}
SU(2)_0 \times SU(2)_0+\psi^{\text{bifund}} \quad\xrightarrow{m_\psi \simeq 0}\quad K_{CS}= \begin{pmatrix} 1&-1\\-1&1 \end{pmatrix}.
\end{align}

It turns out that $\begin{pmatrix} 1&-1\\-1&1 \end{pmatrix}\simeq U(1)_1 \times U(1)_0$ under the $SL(2,\mathbb Z)$ transformation (see \cite{Delmastro:2019vnj} for a recent discussion) and, since $U(1)_1$ is trivial, $U(1)_0$ is isomorphic to a compact boson. We see that the intermediate quantum phase is isomorphic to the one obtained in the $Spin(4)+2~\psi^{\text{vec}}$ theory.

It is quite interesting to see the global symmetry matching between these two distinct constructions. On the $Spin(4)$ side, the $SO(2)$ flavor symmetry rotating two vector fermions lead to a  nonlinear $\sigma$-model with $S^1$. The scalar dual description of it preserves the $SO(2)$ flavor symmetry and the symmetry breaking is realized through the condensation of the scalar. On the $SU(2)\times SU(2)$ side, the $SO(2)$ flavor symmetry is translated to the $U(1)$ baryon symmetry, and either of the scalar or fermion dual descriptions have a $U(1)$ monopole symmetry which doesn't act on the matter field. When we deform the mass of the dual description to flow into the quantum phase, the monopole symmetry in the ultraviolet flows to the $U(1)$ shift symmetry of the compact boson, where the symmetry breaking is triggered by the non-trivial monopole flux.

\section{Generalized Fermion Particle-Vortex Duality}
\label{app:ferm_pv}

In this appendix, we elaborate on how the flavor-bounded quiver we constructed in Sec. \ref{sec:flavor-bounded} is a generalization the fermionic particle-vortex duality proposed by Son \cite{Son:2015xqa}. As a reminder, this duality also lives in $2+1$-dimensions and conjectures a free Dirac fermion is dual to a $U(1)\times U(1)$ Chern-Simons matter theory, which we refer to as ``QED$_3$''. The generalization is a straightforward extension of previous work \cite{Aitken:2018cvh}, where it was shown that the bosonic two-node quiver duality can be viewed as a generalization of the bosonic particle-vortex duality. We begin by emphasizing the special case which reduces to Son's duality, which corresponds to taking $N_1 = N_2 = k_1 = k_2=1$.

\subsection{Son's Particle-Vortex Duality}

Let us begin by reviewing how Son's duality is constructed from the Abelian limits of \eqref{eq:aharony} as it was done in Ref. \cite{Seiberg:2016gmd}. We will then show this is simply the Abelian limit of the two-node quiver construction we performed above. The ``scalar + flux = fermion'' duality is the $N=k=N_f=1$ limit of \eqref{eq:aharony 1}, which at the Lagrangian level is
\begin{subequations}
\begin{align}
\mathcal{L}_{SU} & =i\bar{\psi}\slashed{D}_{\tilde{A}_{1}}\psi \label{eq:free_ferm} \\
\mathcal{L}_{U} & =\left|D_{c}\Phi\right|^{2}-i\left[\frac{1}{4\pi}cdc-\frac{1}{2\pi}cd\tilde{A}_{1}\right] \label{eq:son1}
\end{align}
\end{subequations}
with mass mapping $m_{\psi}\leftrightarrow-m_{\Phi}^{2}$. The (time-reversed) ``fermion
+ flux = scalar'' is the same limit of \eqref{eq:aharony 2}, given by
\begin{subequations}
\label{eq:ferm_flux}
\begin{align}
\mathcal{L}_{SU} & =\left|D_{\tilde{A}_{1}}\phi\right|^{2}-i\left[-\frac{1}{4\pi}\tilde{A}_{1}d\tilde{A}_{1}\right]\\
\mathcal{L}_{U} & =i\bar{\Psi}\slashed{D}_{c}\Psi-i\left[\frac{1}{4\pi}cdc-\frac{1}{2\pi}cd\tilde{A}_{1}\right]
\end{align}
\end{subequations}
which has a mass mapping $m_{\Psi}\leftrightarrow m_{\phi}^{2}$.

In the above expressions, we have a free fermion in \eqref{eq:free_ferm}, which automatically gives one side of Son's duality. Our goal should be to match the scalar side, \eqref{eq:son1}, using the fermion + flux duality. We can take \eqref{eq:ferm_flux} and add $-i\left[\frac{2}{4\pi}\tilde{A}_{1}d\tilde{A}_{1}-\frac{1}{2\pi}\tilde{A}_{1}d\tilde{B}_{1}\right]$ to both ends and promote the background gauge field, $\tilde{A}_{1}\to\tilde{a}_{1}$, which yields
\begin{subequations}
\begin{align}
\mathcal{L}_{SU} & =\left|D_{\tilde{a}_{1}}\phi\right|^{2}-i\left[\frac{1}{4\pi}\tilde{a}_{1}d\tilde{a}_{1}-\frac{1}{2\pi}\tilde{a}_{1}d\tilde{B}_{1}\right]\\
\mathcal{L}_{U} & =i\bar{\Psi}\slashed{D}_{c}\Psi-i\left[\frac{1}{4\pi}cdc-\frac{1}{2\pi}cd\tilde{a}_{1}+\frac{2}{4\pi}\tilde{a}_{1}d\tilde{a}_{1}-\frac{1}{2\pi}\tilde{a}_{1}d\tilde{B}_{1}\right].
\end{align}
\end{subequations}
The scalar end of these dualities are identical under the identification $\phi\Leftrightarrow\Phi$, $\tilde{a}_{1}\Leftrightarrow c$, and $\tilde{A}_{1}\Leftrightarrow\tilde{B}_{1}$. Thus we arrive at the duality
\begin{equation}
i\bar{\psi}\slashed{D}_{\tilde{A}_{1}}\psi\qquad\leftrightarrow\qquad i\bar{\Psi}\slashed{D}_{c}\Psi-i\left[\frac{1}{4\pi}cdc-\frac{1}{2\pi}cd\tilde{a}_{1}+\frac{2}{4\pi}\tilde{a}_{1}d\tilde{a}_{1}-\frac{1}{2\pi}\tilde{a}_{1}d\tilde{A}_{1}\right]\label{eq:sons pv}
\end{equation}
with the mass identification $m_{\psi}\leftrightarrow-m_{\Psi}$. Recall, we are using the notation where the fermion comes with a default level $-1/2$, so it may look slightly different than what was is sometimes in the literature. This is the properly quantized form of the duality also found in Ref. \cite{Seiberg:2016gmd}. This construction is essentially the time-reversed version of the derivation performed there. As discussed in Ref. \cite{Seiberg:2016gmd}, integrating out the dynamical gauge field $c$ brings the right-hand side of \eqref{eq:sons pv} to its simpler and more familiar form, at the cost of violating Dirac quantization.

\begin{figure}
\begin{centering}
\includegraphics[scale=0.65]{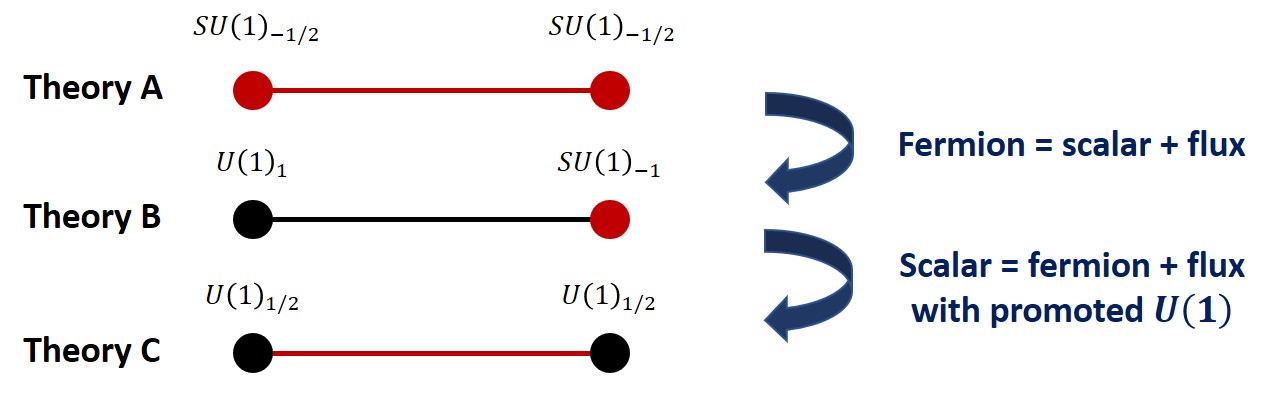}
\par\end{centering}
\caption{Fermion particle-vortex duality as a two-node quiver theory.\label{fig:Fermion-particle-vortex-duality}}
\end{figure}

To match the duality to two-node quivers we shift the gauge field on the right-hand-side of \eqref{eq:sons pv} to $c\to c+\tilde{a}_{1}+\tilde{A}_1$, which then gives
\begin{equation}
\mathcal{L}_{U}=i\bar{\Psi}\slashed{D}_{c+\tilde{a}_{1}+\tilde{A}_{1}}\Psi-i\left[\frac{1}{4\pi}cdc+\frac{1}{4\pi}\tilde{a}_{1}d\tilde{a}_{1}+\frac{1}{4\pi}\tilde{A}_{1}d\tilde{A}_{1}\right].\label{eq:son shifted}
\end{equation}
One can check that, after a relabeling of dynamical gauge fields and rearranging the background Chern-Simons terms, this is simply the $N_1=N_2=k_1=k_2=1$ limit of the Lagrangians in \eqref{eq:flaovr-bounded lag}. The derivation is summarized in Fig. \ref{fig:Fermion-particle-vortex-duality}.

\subsection{Theories Dual to the Free Dirac Fermion}

We have just shown the fermion particle-vortex duality admits generalizations in the form of the flavor-bounded two-node quiver we constructed in Sec. \ref{sec:flavor-bounded}. Generalizations can then be constructed by tuning the three independent parameters: $N_1$, $N_2\;(=k_1)$, and $k_2$.

For instance, there is an infinite class of dualities which continue to just have a fermion on one end of the duality, so long as $N_1=N_2=1$. We can vary $k_2$ and get a theory which would is still dual to the free Dirac fermion,
\begin{subequations}
\begin{align}
\mathcal{L}_{SU} & =i\bar{\psi}\slashed{D}_{\tilde{A}_{1}}\psi-i\left[\frac{1}{4\pi}\tilde{A}_{1}d\tilde{A}_{1}\right],\\
\mathcal{L}_{U} & =i\bar{\Psi}\slashed{D}_{c+g+\tilde{A}_{1}}\Psi-i\left[\frac{1}{4\pi}gdg+\frac{1}{4\pi}\text{Tr}_{k_{2}}\left(cdc-i\frac{2}{3}c^{3}\right)\right],
\end{align}
\end{subequations}
where now $c\in U(k_2)$. On the $SU$ side changing $k_2$ only effects the level of the non-existent Chern-Simons term. This also opens up the possibility of taking $k_{2}$ to be large and possibility doing some sort of $1/k_{2}$ expansion \cite{Hui:2017cyz}.

\subsection{Theories Similar to ``$\text{QED}_{3}$''}

Also interesting is the case of $N_{2}=k_{1}=k_{2}=1$ with $N_{1}$ kept arbitrary, where we still get two Abelian fields on the $U$ side of the duality so the theory is similar to $\text{QED}_{3}$. However, what will change is the relative levels of the Chern-Simons terms on this end of the duality. This would be dual to a fermion coupled only to an $SU$ field. Again, it would be interesting to take $N_{1}$ to be large and do some $1/N_{1}$ expansion. Explicitly,
\begin{subequations}
\begin{align}
\mathcal{L}_{SU} & =i\bar{\psi}\slashed{D}_{b^{\prime}+\tilde{A}_{1}}\psi-i\left[\frac{N_{1}}{4\pi}\tilde{A}_{1}d\tilde{A}_{1}\right],\\
\mathcal{L}_{U} & =i\bar{\Psi}\slashed{D}_{c+g+\tilde{A}_{1}}\Psi-i\left[\frac{N_{1}}{4\pi}gdg+\frac{1}{4\pi}cdc\right].
\end{align}
\end{subequations}
First off, note that the presence of the level-$1/2$ $b^{\prime}$ Chern-Simons term on the $\psi$ side means this cannot be particle-hole symmetric. That's okay if we're not attempting to describe $\nu=1/2$.

We can make this look more similar to the properly-quantized version of Son's duality by shifting $c\to c-g+\tilde{A}_1$, which makes the $\mathcal{L}_{\Psi}$ becomes
\begin{equation}
\mathcal{L}_{U}=i\bar{\Psi}\slashed{D}_{c}\Psi-i\left[\frac{N_{1}+1}{4\pi}gdg-\frac{1}{2\pi}gdc+\frac{1}{4\pi}cdc-\frac{N_{1}}{2\pi}gd\tilde{A}_{1}+\frac{N_1}{4\pi}\tilde{A}_1 d \tilde{A}_1\right].
\end{equation}
This is the same as $\text{QED}_{3}$ (with a relabeling of $g\to\tilde{a}_{1}$), except now the $g$ and $\tilde{A}_{1}$ Chern-Simons terms have an arbitrary integer level. This reduces to Son's duality in the $N_{1}=1$ limit.

\section{Two-Node Fermionic Quivers for the $SO/Sp$ gauge group} \label{app:so/sp}

Here we generalize the two-node fermionic quiver dualities analyzed in Sec. \ref{sec:two-node} to the case of $SO$ and $Sp$ gauge groups. We only list the main results since the approach is largely similar to the unitary case.

First, we write down the flavor bounded two-node quiver dualties for the $SO/Sp$ gauge group. The construction parallels that of Sec. \ref{sec:flavor-bounded}, but instead using the version of 3d bosonization dualities for the $SO/Sp$ gauge group appeared in Ref. \cite{Aharony:2016jvv}. First, the boson-fermion duality is obtained from 3d bosonization by gauging the full flavor symmetry subject to the flavor-bound $k_1\geq N_2$,
\begin{subequations}
\begin{align}
~SO(N_1)_{-k_1+\frac{N_2}{2}} \times SO(N_2)_{-k_2+\frac{N_1}{2} }+\psi & \quad\leftrightarrow\quad SO(k_1)_{N_1}\times SO(N_2)_{-k_2}+\Phi, \\
Sp(N_1)_{-k_1+\frac{N_2}{2}} \times Sp(N_2)_{-k_2+\frac{N_1}{2} }+\psi & \quad\leftrightarrow\quad Sp(k_1)_{N_1}\times Sp(N_2)_{-k_2}+\Phi.
\end{align}
\end{subequations}
From the above dualities, there exists fermion-fermion dualities similar to \eqref{eq:flaovr-bounded lag} for the special case of $k_1=N_2$,
\begin{subequations}
\begin{align}
SO(N_1)_{-\frac{N_2}{2}} \times SO(N_2)_{-k_2+\frac{N_1}{2} }+\psi & \quad\leftrightarrow\quad SO(N_2)_{N_1-\frac{k_2}{2}}\times SO(k_2)_{\frac{N_2}{2}}+\Psi, \\
Sp(N_1)_{-\frac{N_2}{2}} \times Sp(N_2)_{-k_2+\frac{N_1}{2} }+\psi & \quad\leftrightarrow\quad Sp(N_2)_{N_1-\frac{k_2}{2}}\times Sp(k_2)_{\frac{N_2}{2}}+\Psi.
\end{align}
\end{subequations}

The flavor-violated version of boson-fermion two-node quiver dualities for the $SO/Sp$ gauge group are built from the extension of bosonization dualities for the $SO/Sp$ gauge group beyond the flavor bound analyzed in Ref. \cite{Komargodski:2017keh}. The results are similar to \eqref{eq:2nodebosefermi} and valid for $k_1<N_2,~k_2<N_1$,
% \footnote{We again remark that the bosonic dual descriptions are natural since these are obtained directly from the gauging of the flavor symetry in the dualities of \cite{Komargodski:2017keh}.}
\begin{equation}
\begin{aligned}
SO(N_1)_{-k_1+N_2/2}\times &SO(N_2)_{-k_2+N_1/2} +\psi
\\ &\leftrightarrow\quad\begin{cases}
SO(k_1)_{N_1}\times SO(N_2)_{-k_2}+\phi_1& m_{\psi}=-m_{*},\\
SO(N_2-k_1)_{-N_1}\times SO(N_2)_{N_1-k_2}+\phi_2 & m_{\psi}=m_{*}\end{cases}
\\Sp(N_1)_{-k_1+N_2/2}\times &Sp(N_2)_{-k_2+N_1/2} +\psi
\\ &\leftrightarrow\quad\begin{cases}
Sp(k_1)_{N_1}\times Sp(N_2)_{-k_2}+\phi_1& m_{\psi}=-m_{*}\\
Sp(N_2-k_1)_{-N_1}\times Sp(N_2)_{N_1-k_2}+\phi_2 & m_{\psi}=m_{*}.
\end{cases}
\end{aligned}
\end{equation}

Finally, the new fermionic two-node quiver dual descriptions, which equivalently describe the above phase transitions, for the $SO/Sp$ cases are
\begin{equation}
\begin{aligned}
SO(N_1)_{-k_1+N_2/2}& \times SO(N_2)_{-k_2+N_1/2} +\psi
\\ \leftrightarrow\quad & \begin{cases}
SO(k_1)_{N_1-k_2/2}\times SO(k_2)_{N_2-k_1/2}+\Psi_1& m_{\psi}=-m_{*}\\
SO(N_1-k_2)_{-N_2/2-k_1/2}\times SO(N_2-k_2)_{-N_1/2-k_2/2}+\Psi_2 & m_{\psi}=m_{*},\end{cases}
\\Sp(N_1)_{-k_1+N_2/2}& \times Sp(N_2)_{-k_2+N_1/2} +\psi
\\ \leftrightarrow\quad & \begin{cases}
Sp(k_1)_{N_1-k_2/2}\times Sp(k_2)_{N_2-k_1/2}+\Psi_1& m_{\psi}=-m_{*}\\
Sp(N_1-k_2)_{-N_2/2-k_1/2}\times Sp(N_2-k_2)_{-N_1/2-k_2/2}+\Psi_2 & m_{\psi}=m_{*}.\end{cases}
\end{aligned}
\end{equation}

We comment that $SO/Sp$ case is simpler than the unitary case since the corresponding level/rank duality preserves the type of Lie group. Similar subtleties as the unitary case will arise if one constructs the two-node quiver for the various modification of the orthogonal or symplectic group with discrete gauging/extension (e.g. $Spin, Pin^\pm,...$), as analyzed for the orthogonal case in Ref. \cite{Cordova:2017vab}.

\section{Gravitational Counterterm Matching}
\label{app:gravitational}

Here we show that the gravitational counterterms are consistent along the phase diagram of the two-node quiver discussed in Sec. \ref{sec:two-node}. In our convention, we define the gravitational counterterm as the coeffcient of twice the gravitational Chern-Simons term $2\text{CS}_\text{grav}$, where $\int_{M=\partial X}\text{CS}_\text{grav}=\frac{1}{192\pi}\int_X\text{tr}R \wedge R$. Refs. \cite{Benini:2017aed, Aitken:2018joi} contain many of the 3d bosonization dualities used throughout this work with gravitational Chern-Simons terms made explicit.

First, let's discuss the boson-fermion dualities between bifundamental matter in \eqref{eq:2nodebosefermi}. Consistency of the phase diagram requires that following a non-trivial closed path in phase space should have net zero change of the gravitational counterterm
\begin{equation} \label{eq:grav}
\begin{aligned}
\Delta c \left[(\text{A I}) \rightarrow  (\text{B$_1$ I}) \rightarrow  (\text{B$_1$ III})
 \rightarrow (\text{B$_2$ III}) \rightarrow (\text{B$_2$ II}) \rightarrow  (\text{A II}) \rightarrow (\text{A I}) \right]=0
\end{aligned}
\end{equation}
where $(\text{A I})$ represents phase $\text{I}$ of Theory A, etc. It turns out that there are two kinds of contributions to the gravitational counterterm along the path above. First, there is a contribution coming from the difference of the one-loop determinant of the fermion coupled to the background metric between negative or positive mass, which is equivalent to the complex dimension of the representation. Second, there is one coming from the compensating gravitational Chern-Simons term along the level-rank duality in the various phases, which is explained in Ref. \cite{Hsin:2016blu}. Namely, $\Delta c=-Nk$ along the level-rank duality from $SU(N)_k$ to $U(k)_{-N}$. It is important to point out that scalar doesn't contribute to the gravitational counterterm and the middle phases described by the two scalar dual descriptions $\text{B}_1$ and $\text{B}_2$ are equivalent without the need of any duality transformation. After tracking down the non-trivial closed path in \eqref{eq:grav}, one can show that the gravitational counterterm is consistent, i.e. $\Delta c \vert_{\text{closed path}}=0$.

Now the above test could also be done with the two-node fermionic dual description described in \eqref{eq:2nodefermifermi}. Using a similar procedure and formulas as above, consistency of the gravitational counterterms can be established.

\end{appendix}

\bibliographystyle{JHEP}
\bibliography{fermionquivers}
\end{document}